# Buckling and post-buckling of cylindrical shells under combined torsional and axial loads

Lu Lu, Sophie Leanza, Yang Liu, Ruike Renee Zhao*

*Department of Mechanical Engineering, Stanford University, Stanford, CA 94305, United States*

*Corresponding author: rrzhao@stanford.edu

## Abstract

The buckling behavior of cylindrical shells has gained significant interest over the past century due to its rich nonlinear behavior and broad engineering applications. While the buckling of cylindrical shells under a single load (e.g., compression or torsion) has been extensively studied, the buckling behavior under combined torsional and axial loads remains largely unexplored. In this paper, based on a combination of experiments, theoretical modeling, and finite element simulations, we systematically investigate the buckling and post-buckling behavior of cylindrical shells under combined torsional and axial loads. Three different types of combined loads are considered: compression with pre-torsion, torsion with pre-tension, and torsion with pre-compression. The theoretical model is established within the framework of the Donnell shell theory and solved using the Galerkin method, through which the critical buckling load, critical circumferential wavenumber, buckling pattern, and post-buckling equilibrium path of clamped-clamped thin cylindrical shells under various types of loads can be determined. The theoretical predictions agree well with finite element simulations and qualitatively capture the various buckling phenomena observed in the experiments. It is found that cylindrical shells exhibit quite different post-buckling behavior under combined loads compared to under a single compressive or torsional load. For instance, when a clamped-clamped thin cylindrical shell is subjected to pure torsion or torsion with a relatively small pre-compression, it consistently shows a diagonal-shaped pattern during deformation. However, with a relatively large pre-compression, the shell transitions from a diagonal-shaped pattern to a twisted diamond-shaped pattern. Our work reveals the role of torsion-compression/tension coupling in the buckling instabilities of cylindrical shells, which can guide the design of cylindrical shell buckling-inspired foldable structures such as origami systems driven by combined loads.

# 1. Introduction

Buckling of cylindrical shells has long attracted considerable research attention in the mechanics community due to its rich fundamental insights into instability and its engineering importance in evaluating the structural safety of shell-based structures. In the past century, cylindrical shell buckling problems, including critical buckling load (Batdorf et al., 1947; Donnell, 1935; Timoshenko and Gere, 1961; Weingarten et al., 1968), post-buckling path evolution (Budiansky, 1967; Hunt and Neto, 1991; Hutchinson and Koiter, 1970; Sun et al., 2020; Tvergaard, 1983; Von Karman and Tsien, 1941; Yamaki, 1984), imperfection sensitivity (Hutchinson, 2010; Hutchinson and Thompson, 2018; Simitses, 1986; Zhang and Han, 2007), and wrinkling morphology (Cao et al., 2012; Xu and Potier-Ferry, 2016; Yang et al., 2018; Zhao et al., 2014), have been extensively studied. It is well-understood that cylindrical shells lose stability and buckle into an axisymmetric or diamond-shaped pattern when subjected to a sufficient axial compressive load (Horton and Durham, 1965), and deform into a diagonal-shaped pattern when subjected to a sufficient torsional load (Hunt and Ario, 2005). These two buckling phenomena inspired the discovery of the well-known Yoshimura and Kresling origami patterns (Kresling, 2008; Yoshimura, 1955), which have found widespread applications in soft robotics (Melancon et al., 2022; Wu et al., 2021; Ze et al., 2022a; Ze et al., 2022b; Zhang et al., 2023) and mechanical metamaterials (Wang et al., 2023; Yasuda et al., 2019; Zhai et al., 2018; Zhang and Rudykh, 2024). In recent years, benefitting from its rich nonlinear behavior, cylindrical shell buckling has also become a promising platform for designing various functional structures, such as wrinkled cylindrical sheet metamaterials with tunable mechanical response (Dong et al., 2023), nonuniform thickness soft cylindrical shells capable of multimodal deformation (Yang et al., 2024), and auxetic meta-shells that can suppress torsional instability under twisting (Ghorbani et al., 2024).

To date, cylindrical shell buckling has mainly been studied under a single load, either pure compression or pure torsion, with little attention paid to the buckling behavior under combined torsional and axial loads. Elastic structures under combined loads can exhibit much richer buckling behavior compared to those under a single load. For example, adjusting the twist angle and tension of a twisted flat elastic ribbon can vary the buckling modes between a helicoid, a longitudinally/transversely buckled helicoid, a creased helicoid, and a localized loop (Chopin and Kudrolli, 2013; Liu et al., 2024). A bent elastic ribbon can trigger snap-through instability under twisting (Huang et al., 2024; Sano and Wada, 2019). For cylindrical shells, early studies have indicated that the axial force has significant influence on the critical buckling load under torsion (Batdorf et al., 1947; Bisagni and Cordisco, 2003; Meyer-Piening et al., 2001; Winterstetter and Schmidt, 2002). Specifically, tension tends to increase the critical torsional buckling load, while compression reduces the critical torsional buckling load (Batdorf et al., 1947; Timoshenko and Gere, 1961). Nevertheless, compared to the widely studied post-buckling behavior of cylindrical shells under pure compression or pure torsion, the post-buckling behavior under combined torsional and axial loads remains largely unexplored. For example, the post-buckling equilibrium path and the post-buckling pattern evolution under combined torsional and axial loads have not been previously studied. Filling this gap could offer valuable insights into the role of torsion-compression/tension coupling in the buckling instabilities of cylindrical shells, providing guidelines for the design of functional cylindrical shell structures or cylindrical shell buckling-inspired origami systems enabled by combined loads.

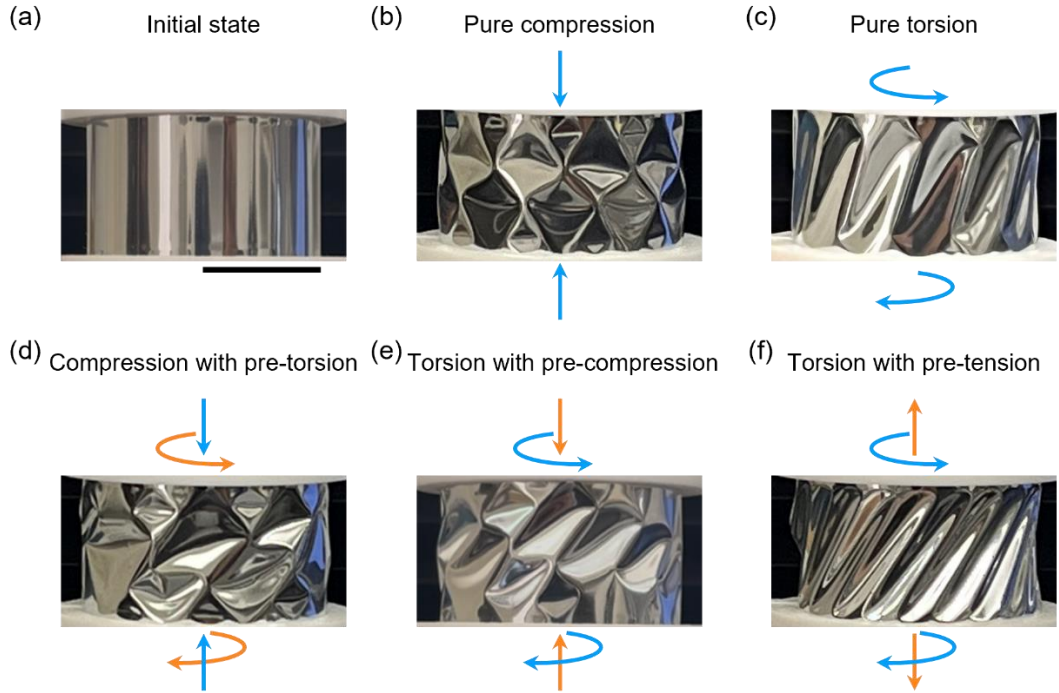

**Fig. 1.** Experimental images of buckling patterns of a cylindrical shell under a single load or combined torsional and axial loads. (a) Initial state of the cylindrical shell; buckling pattern under (b) pure compression; (c) pure torsion; (d) compression with pre-torsion; (e) torsion with pre-compression; (f) torsion with pre-tension. The orange arrows represent the preloads. Scale bar: 20 mm.

**Fig. 1** presents preliminary experimental images of the buckling patterns of a cylindrical shell (**Fig. 1(a)**) under a single load or combined torsional and axial loads (see **Video 1** in the Supplementary Materials). It is shown that the shell exhibits the diamond-shaped and diagonal-shaped buckling patterns under pure compression (**Fig. 1(b)**) and pure torsion (**Fig. 1(c)**), respectively. However, when the cylindrical shell is subjected to compression with pre-torsion (**Fig. 1(d)**), a twisted diamond-shaped pattern is obtained, highlighting the significant influence of pre-torsion on the buckling pattern of cylindrical shells under compression. Additionally, when pre-compression is combined with torsion (**Fig. 1(e)**), the shell also shows a twisted diamond-shaped pattern. In contrast, when pre-tension is combined with torsion (**Fig. 1(f)**), the shell exhibits a diagonal-shaped pattern with a significantly increased circumferential wavenumber compared to the pure torsion case. This demonstrates that the axial load can tune the buckling pattern as well as the circumferential wavenumber of cylindrical shells under torsion.

To fully understand the buckling and post-buckling behavior of cylindrical shells under combined torsional and axial loads, in this work, we systematically investigate this problem through a combination of experiments, theoretical modeling, and finite element simulations. Three types of combined loads are considered, compression with pre-torsion, torsion with pre-tension, and torsion with pre-compression. The theoretical framework for the buckling and post-buckling analysis is established based on the Donnell shell theory and then solved using the Galerkin method. Finite element analyses (FEA) are performed to compare with the theoretical solutions, and a good agreement between them is achieved. Based on the buckling analysis, we obtained the phase diagram for the critical buckling load, critical circumferential wavenumber, and critical buckling pattern of clamped-clamped (C-C) thin cylindrical shells under combined torsion and compression. From the post-buckling analysis, we determined the post-buckling equilibrium paths and the pattern evolution rules of C-C thin cylindrical shells under compression with pre-torsion or under torsion with pre-tension/compression. The obtained results qualitatively capture the various buckling phenomena observed in the experiments. Our work reveals the role of torsion-compression/tension coupling in the buckling instabilities of cylindrical shells, which could provide guidelines for the design of shell structures under combined loads. Also, we envision that the various buckling patterns presented in this work can inspire the discovery of new origami patterns to enable more functional foldable structures under combined loads.

The remainder of this paper is organized as follows. In **Section 2**, we introduce the experimental results of post-buckling behavior of a C-C cylindrical shell under various types of loads. In **Section 3**, we first review the governing equations for shell buckling analysis based on the Donnell shell theory and then solve the critical buckling and post-buckling problems under combined torsional and axial loads using the Galerkin method. In **Section 4**, FEA details for our shell buckling simulations are provided. In **Section 5**, the results of buckling and post-buckling of C-C thin cylindrical shells obtained by the theoretical model and FEA simulations are both presented. In **Section 6**, we summarize the main conclusions of this work.

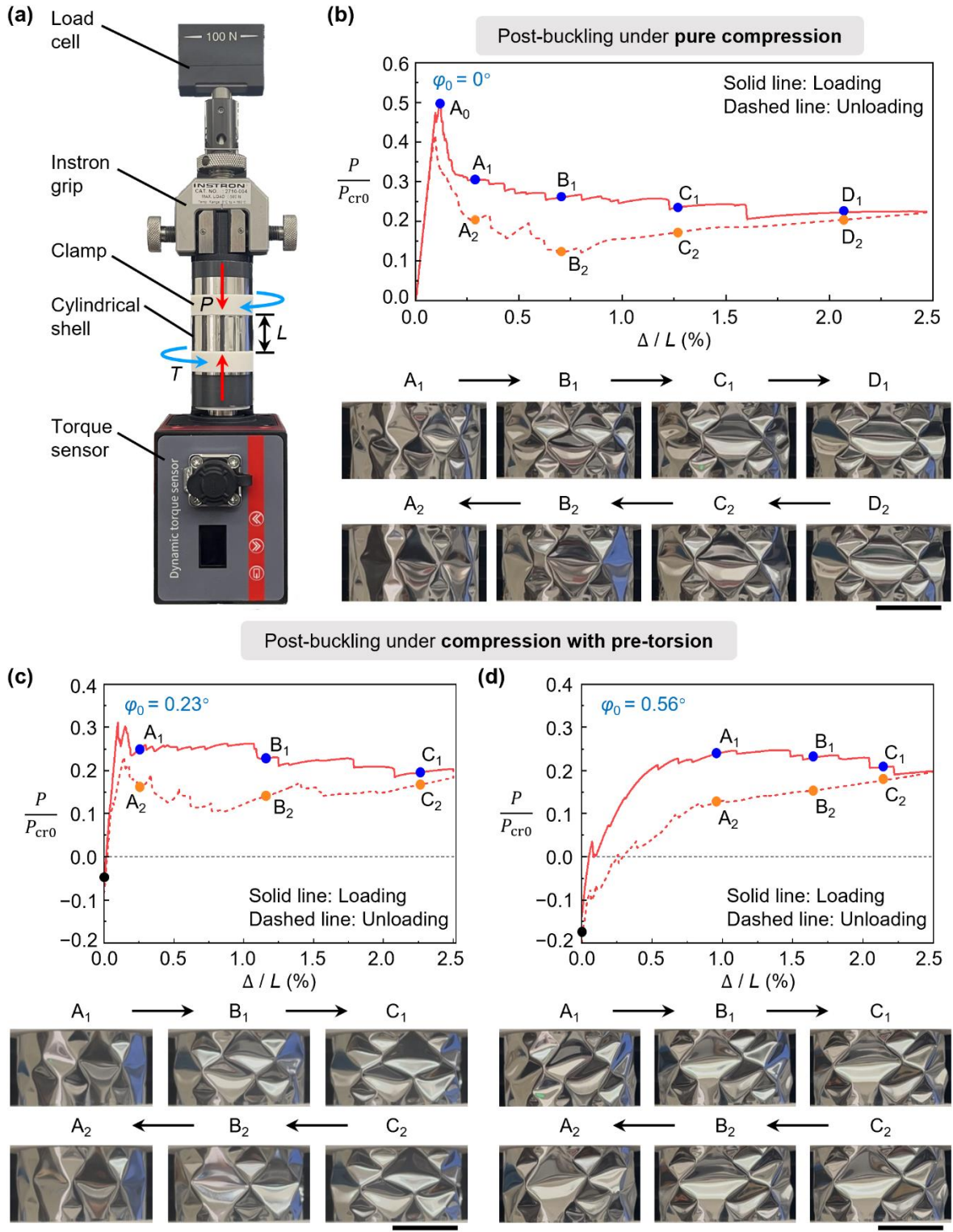

**Fig. 2.** Experimental results of the normalized compressive force versus the normalized displacement for post-buckling behavior of a C-C cylindrical shell under pure compression or compression with pre-torsion. (a) Experimental setup for the post-buckling tests under a single load or combined loads. (b) Post-buckling

equilibrium path of a C-C cylindrical shell under pure compression. Here, $P$ is the compressive force, $P_{cr0}$ is the classical critical buckling load of cylindrical shells with simply supported ends given in Eq. (1), $\Delta$ is the displacement (i.e., end shortening), and $L$ is the initial length of the cylindrical shell. (c) Post-buckling equilibrium path of a C-C cylindrical shell under compression with a relatively small pre-torsion. The pre-torsion is applied by twisting the shell's end to a pre-twist angle of $\varphi_0 = 0.23°$, which is then held constant throughout the loading and unloading processes. (d) Post-buckling equilibrium path of a C-C cylindrical shell under compression with a relatively large pre-torsion. The pre-torsion is applied by twisting the shell's end to a pre-twist angle of $\varphi_0 = 0.56°$, which is then held constant throughout the loading and unloading processes. In Figs. (c) and (d), the measured compressive force begins from a negative value due to a small tension induced by the pre-twist in the initial state, and the black dot denotes the starting point of the compression test. Scale bars: 20 mm.

## 2. Experiments

We first conduct experiments to investigate how the post-buckling behavior of cylindrical shells is affected by combined loads, including compression with pre-torsion, torsion with pre-tension, and torsion with pre-compression. For comparison purposes, post-buckling under pure compression or pure torsion is also investigated. The experiments are carried out on an Instron universal testing machine with a mounted torque sensor, as shown in **Fig. 2(a)**. The tested cylindrical shell is fabricated by wrapping a thin mylar film (without gluing) around the solid cylinders connected to the Instron and torque sensor and fixing its two ends within the two clamps. The distance between the clamps is considered to be the initial length of the cylindrical shell, where $L = 20$ mm. The thin film has a uniform thickness of $h = 25$ μm, and the fabricated shell has a uniform radius of $R = 19.05$ mm. Young's modulus $E$ and Poisson's ratio $v$ of the mylar film are 3.5 GPa and 0.3, respectively.

During loading, compression/tension or torsion is applied by translating or rotating one of the solid cylinders. Specifically, the upper solid cylinder connected to the load cell of Instron moves along the cylinder's longitudinal direction for compression/tension loading, while the lower solid cylinder attached to the torque sensor rotates for torsion loading. For compression tests, a prescribed displacement is applied at the top end, and a specified pre-twist angle $\varphi_0$ between both ends is held constant throughout the test for cases with pre-torsion. Due to the applied pre-twist, no additional rotation is allowed at either end of the shell during compression tests. In the pure torsion test, the top end is free to translate in the longitudinal direction while a prescribed twisting angle is applied, and the corresponding torque is measured at the bottom end. For torsion with pre-tension/compression tests, a specified axial preload $P_0$ is held constant throughout the test.

**Fig. 2(b)** shows the normalized force-displacement curve of the C-C cylindrical shell under pure compression (i.e., pre-twist $\varphi_0 = 0°$), where the displacement $\Delta$ is normalized by the shell's initial length, and the compressive force $P$ is normalized by the classical critical buckling load, referred to as the classical value, of a cylindrical shell with simply supported ends under pure compression obtained using thin shell theory (Timoshenko and Gere, 1961),

$$P_{cr0} = \frac{2\pi E h^2}{\sqrt{3(1-v^2)}}. \tag{1}$$

Note that the normalized compressive force $P/P_{cr0}$ reflects how much the compressive force is in the post-buckling regime compared to the critical buckling load under pure compression. This parameter will also be used in our numerical results for post-buckling under compression, as presented in **Section** 4.2. It is seen that the compressive force linearly increases with the displacement until reaching the critical buckling state. However, the measured critical buckling load $P_{cr} = 4.20$ N (corresponding to the point $A_0$ in **Fig. 2(b)**) is only about 50% of the classical value ($P_{cr0} = 8.32$ N for the shell used in our experiments), showing that the critical buckling load is sensitive to geometric imperfections (Hutchinson and Thompson, 2018; Simitses, 1986). In our experiments, minor dimple-like geometric imperfections may be introduced during fabrication. After the onset of buckling, a series of snapping processes occurs and the shell snaps from one buckling mode to another with a decreased buckling load. During each snapping process, the circumferential wavenumber decreases by one, while the axial wavenumber remains unchanged. The corresponding buckling pattern evolution in the loading process is shown at the bottom of **Fig. 2(b)** and in **Video 2** in the Supplementary Materials. We can observe that the shell exhibits a diamond-shaped pattern under pure compression. Upon unloading, the deformation follows a new equilibrium path with lower buckling load, during which the circumferential wavenumber remains unchanged while the axial wavenumber reduces from 2 to 1. This indicates that the shell shows an irreversible loading/unloading equilibrium path under pure compression.

When the shell is subjected to compression with a pre-torsion, the measured post-buckling equilibrium paths are illustrated in **Fig. 2(c)** and **(d),** and the corresponding experimental loading processes are provided in **Video 3** in the Supplementary Materials. As mentioned before, the pre-torsion is applied by twisting the shell's end to a pre-twist angle $\varphi_0$, which is then held constant throughout the loading and unloading processes. For the two cases considered in **Fig. 2(c)** and **(d)**, $\varphi_0 = 0.23°$ and $0.56°$, respectively. Note that the measured compressive force begins from a negative value due to a small tension induced by the pre-twist in the initial state, and the starting point of the compression test is denoted by a black dot on the curve. When the pre-twist is relatively small (e.g., $\varphi_0 = 0.23°$ in **Fig. 2(c)**), the shell shows a similar post-buckling behavior to the pure compression case and exhibits a diamond-shaped pattern during loading and unloading. However, when the pre-twist is relatively large (e.g., $\varphi_0 = 0.56°$ in **Fig. 2(d)**), the compressive force initially increases and reaches a plateau, then gradually decreases as snapping occurs, during which the shell shows a twisted diamond-shaped pattern. Therefore, pre-torsion can significantly influence the post-buckling equilibrium paths as well as the buckling patterns of cylindrical shells under compression.

Experimental results of post-buckling behavior of the C-C cylindrical shell under pure torsion (i.e., axial preload $P_0 = 0$) are shown in **Fig. 3(a)** and **Video 4** in the Supplementary Materials. Here, due to the lack of a classical value for the critical buckling torque, the measured torque $T$ is nondimensionalized by the flexural rigidity $D = Eh^3/[12(1-v^2)]$ and geometric parameters $R$ and $L$ of the shell as: $k_s = TL^2/(2\pi^3 R^2 D)$ (Yamaki, 1984). This dimensionless torque will also be used to present our numerical results for post-buckling under torsion, as presented in **Section** 4.3. It can be observed that the torque linearly increases until critical buckling occurs, after which it gradually decreases. Throughout the entire loading process, no snapping occurs, and the shell exhibits a diagonal-shaped pattern. When the shell is subjected to torsion with a pre-tension $P_0$ (the magnitude of the applied pre-tension is 10 N, which is about 120% of the classical value $P_{cr0} = 8.32$ N under pure compression, and it is held roughly constant during the loading process), the measured torque continuously increases after critical buckling occurs, and the shell exhibits a diagonal-shaped pattern similar to that under pure torsion, but with an increased circumferential wavenumber (see **Fig. 3(b)** and **Video 4** in the Supplementary Materials). Moreover, when the shell is subjected to torsion with a small pre-compression (e.g., $P_0 = 0.12P_{cr0}$ in **Fig. 3(c)** and **Video 5** in the Supplementary Materials), the shell exhibits a similar post-buckling behavior to the pure torsion case: the torque gradually decreases with the twisting angle after buckling. However, when the applied pre-compression is slightly larger (e.g., $P_0 = 0.26P_{cr0}$ in **Fig. 3(d)** and **Video 5** in the Supplementary Materials), the measured torque decreases to zero and then becomes negative after the onset of buckling, and the shell transitions from a diagonal-shaped pattern into a twisted diamond-shaped pattern. These observations demonstrate that the cylindrical shell exhibits a much richer post-buckling behavior under combined torsional and axial loads. In this paper, to fully understand the role of combined loads in the post-buckling behavior of cylindrical shells, we develop a theoretical framework and utilize FEA simulations to investigate the problem.

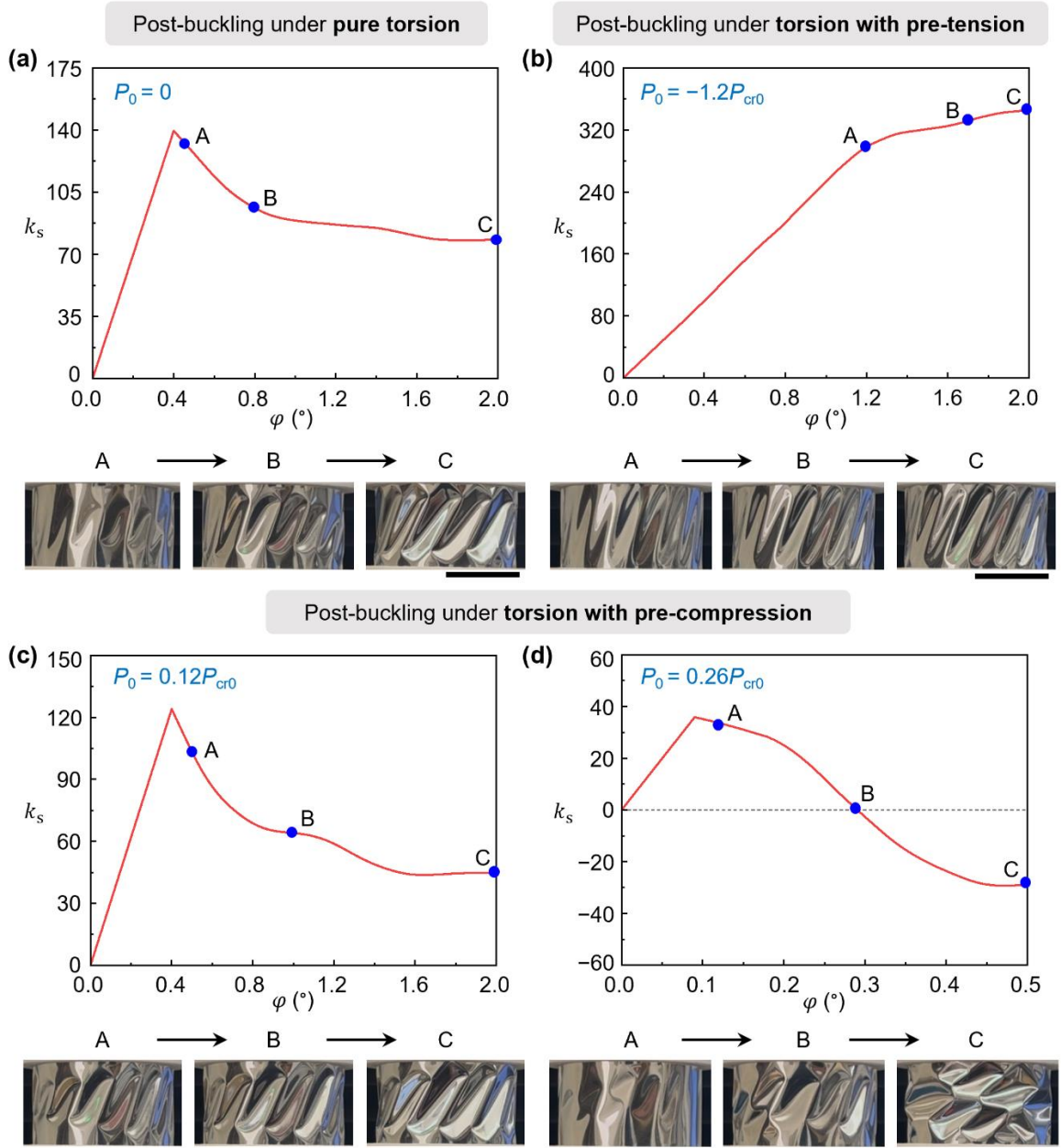

**Fig. 3.** Experimental results of the dimensionless torque versus the twisting angle for post-buckling behavior of a C-C cylindrical shell under (a) pure torsion, (b) torsion with pre-tension $P_0 = -1.2P_{cr0}$, (c) torsion with pre-compression $P_0 = 0.12P_{cr0}$, and (d) torsion with pre-compression $P_0 = 0.26P_{cr0}$. Here, $k_s = TL^2/(2\pi^3 RD)$ is the dimensionless torque with $L$, $R$, and $D$ being the length, radius, and flexural rigidity of the shell, respectively, and $\varphi$ is the twisting angle.

## 3. Theoretical modeling

In this section, the Donnell shell theory (Donnell, 1935) is employed to model the buckling and post-buckling behavior of thin cylindrical shells under combined torsional and axial loads. Due to its relative simplicity and practical accuracy, the Donnell shell theory has been widely used to study the mechanical behavior of thin shell structures. Here, to avoid repetition, we directly present the governing equations of the Donnell shell theory. The derivation details are provided in the **Appendix** for completeness.

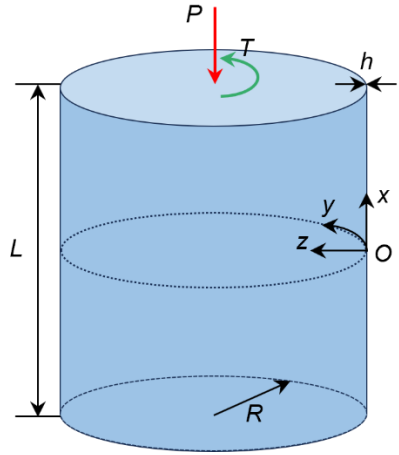

**Fig. 4.** Schematic of a cylindrical shell under combined torsional and axial loads.

*3.1. Governing equations*

As shown in **Fig. 4**, consider a thin elastic cylindrical shell of length $L$, radius $R$, and thickness $h$. The shell is clamped at both ends and subjected to an axial force $P$ and a torque $T$. Here, to keep consistent with most existing literature on shell buckling under compression (Batdorf et al., 1947; Sun et al., 2020; Yamaki, 1984), $P > 0$ is defined to represent compression, and $P < 0$ denotes tension. The two loads are uniformly applied along the edges of the shell, inducing an axial stress $\sigma = P/(2\pi R h)$ and a shear stress $\tau = T/(2\pi R^2 h)$. The coordinate system $(x, y, z)$ is established in the middle of the shell, with the $x$, $y$, $z$ axes along the longitudinal, circumferential, and radial directions, respectively. Based on the Donnell shell theory, the nonlinear governing equations considering large deformations are written as

$$D\nabla^4 w - \frac{1}{R}\frac{\partial^2 F}{\partial x^2} = \frac{\partial^2 F}{\partial y^2}\frac{\partial^2 w}{\partial x^2} - 2\frac{\partial^2 F}{\partial x \partial y}\frac{\partial^2 w}{\partial x \partial y} + \frac{\partial^2 F}{\partial x^2}\frac{\partial^2 w}{\partial y^2}, \tag{2}$$

$$\nabla^4 F + \frac{Eh}{R}\frac{\partial^2 w}{\partial x^2} = Eh\left[\left(\frac{\partial^2 w}{\partial x \partial y}\right)^2 - \frac{\partial^2 w}{\partial x^2}\frac{\partial^2 w}{\partial y^2}\right], \tag{3}$$

with

$$\nabla^4(\bullet) = \frac{\partial^4(\bullet)}{\partial x^4} + 2\frac{\partial^4(\bullet)}{\partial x^2 \partial y^2} + \frac{\partial^4(\bullet)}{\partial y^4}, \quad D = \frac{Eh^3}{12(1-\nu^2)}, \tag{4}$$

$$N_x = \frac{\partial^2 F}{\partial y^2}, \quad N_y = \frac{\partial^2 F}{\partial x^2}, \quad N_{xy} = -\frac{\partial^2 F}{\partial x \partial y}, \tag{5}$$

where $D$ is the flexural rigidity of the shell, $w$ is the transverse displacement, $F$ is the Airy's stress function, and $N_x$, $N_y$, and $N_{xy}$ are resultant forces in the shell defined in Eqs. (A5) and (A6).

For clamped ends, the boundary conditions at $x = -L/2$ and $L/2$ are given by (Yamaki, 1984)

$$w = \frac{\partial w}{\partial x} = \frac{\partial^2 F}{\partial x^2} - \nu\frac{\partial^2 F}{\partial y^2} = \frac{\partial^3 F}{\partial x^3} + (2+\nu)\frac{\partial^3 F}{\partial x \partial y^2} = 0, \tag{6}$$

$$\int_0^{2\pi R} N_x dy = -P, \quad R\int_0^{2\pi R} N_{xy} dy = T. \tag{7}$$

Moreover, the end shortening can be calculated by

$$\Delta = -\frac{1}{2\pi R}\int_0^{2\pi R}\int_{-L/2}^{L/2}\left[\frac{1}{Eh}\left(\frac{\partial^2 F}{\partial y^2} - \nu\frac{\partial^2 F}{\partial x^2}\right) - \frac{1}{2}\left(\frac{\partial w}{\partial x}\right)^2\right]dxdy. \tag{8}$$

The twisting angle is written as

$$\varphi = -\frac{1}{2\pi R^2}\int_0^{2\pi R}\int_{-L/2}^{L/2}\left(\frac{2(1+\nu)}{Eh}\frac{\partial^2 F}{\partial x \partial y} + \frac{\partial w}{\partial x}\frac{\partial w}{\partial y}\right)dxdy. \tag{9}$$

*3.2. Nondimensionalization*

The governing equations and boundary conditions are first nondimensionalized by introducing the following dimensionless quantities,

$$\bar{x} = \frac{\pi x}{L}, \bar{y} = \frac{Ny}{R}, \bar{w} = \frac{w}{h}, f = \frac{F}{Eh^3}, c = \frac{1}{12(1-v^2)},$$
$$\alpha = \frac{L^2}{\pi^2 Rh}, \beta = \frac{NL}{\pi R}, k_x = \frac{PL^2}{2\pi^3 ERh^3}, k_s = \frac{TL^2}{2\pi^3 R^2 D}, \quad (10)$$
$$\bar{\delta} = \frac{R\Delta}{Lh}, \bar{\varphi} = \frac{R^2 \varphi}{Lh},$$

where $\bar{x}$, $\bar{y}$, $\bar{w}$, and $f$ are the dimensionless *x*-coordinate, *y*-coordinate, transverse displacement, and Airy's stress function, respectively. Note that $N$ is the circumferential wavenumber in the buckling mode. Moreover, $c$ is a parameter related to the Poisson's ratio $v$, $\alpha$ is a geometric parameter depending on the length-to-thickness ratio *L/R* and radius-to-thickness ratio *R/h* of the shell, $\beta$ is a dimensionless parameter associated with the circumferential wavenumber, and $k_x$, $k_s$, $\bar{\delta}$, and $\bar{\varphi}$ represent the dimensionless axial force, torque, end shortening, and twisting angle, respectively. Note that the dimensionless axial force $k_x$ used in the theoretical model can be related to its normalized counterpart by $P/P_{cr0} = \sqrt{3(1-v^2)} k_x / \alpha$. Using Eq. (10), the nondimensional governing equations can be written as

$$c\bar{\nabla}^4 \bar{w} - \alpha \frac{\partial^2 f}{\partial \bar{x}^2} = \beta^2 \left( \frac{\partial^2 f}{\partial \bar{y}^2} \frac{\partial^2 \bar{w}}{\partial \bar{x}^2} - 2 \frac{\partial^2 \bar{w}}{\partial \bar{x} \partial \bar{y}} \frac{\partial^2 f}{\partial \bar{x} \partial \bar{y}} + \frac{\partial^2 f}{\partial \bar{x}^2} \frac{\partial^2 \bar{w}}{\partial \bar{y}^2} \right), \quad (11)$$

$$\bar{\nabla}^4 f + \alpha \frac{\partial^2 \bar{w}}{\partial \bar{x}^2} = \beta^2 \left[ \left( \frac{\partial^2 \bar{w}}{\partial \bar{x} \partial \bar{y}} \right)^2 - \frac{\partial^2 \bar{w}}{\partial \bar{x}^2} \frac{\partial^2 \bar{w}}{\partial \bar{y}^2} \right], \quad (12)$$

where

$$\bar{\nabla}^4(\bullet) = \frac{\partial^4(\bullet)}{\partial \bar{x}^4} + 2\beta^2 \frac{\partial^4(\bullet)}{\partial \bar{x}^2 \partial \bar{y}^2} + \beta^4 \frac{\partial^4(\bullet)}{\partial \bar{y}^4}. \quad (13)$$

The nondimensional boundary conditions at $\bar{x} = -\pi/2$ and $\pi/2$ are given by

$$\bar{w} = 0, \quad \frac{\partial \bar{w}}{\partial \bar{x}} = 0, \quad (14)$$

$$\frac{\partial^2 f}{\partial \overline{x}^2} - \nu \beta^2 \frac{\partial^2 f}{\partial \overline{y}^2} = 0, \quad \frac{\partial^3 f}{\partial \overline{x}^3} + (2+\nu)\beta^2 \frac{\partial^3 f}{\partial \overline{x} \partial \overline{y}^2} = 0, \tag{15}$$

$$\left[\frac{\partial f}{\partial \overline{y}}\right]_{\overline{y}=0}^{2\pi} = -\frac{2\pi}{\beta^2} k_x, \quad \left[\frac{\partial f}{\partial \overline{x}}\right]_{\overline{y}=0}^{2\pi} = -\frac{2\pi c}{\beta} k_s. \tag{16}$$

Further, the dimensionless end shortening and twisting angle are

$$\overline{\delta} = -\frac{1}{2\pi^2 \alpha} \int_0^{2\pi} \int_{-\pi/2}^{\pi/2} \left[ \beta^2 \frac{\partial^2 f}{\partial \overline{y}^2} - \nu \frac{\partial^2 f}{\partial \overline{x}^2} - \frac{1}{2}\left(\frac{\partial \overline{w}}{\partial \overline{x}}\right)^2 \right] d\overline{x} d\overline{y}, \tag{17}$$

$$\overline{\varphi} = -\frac{\beta}{2\pi^2 \alpha} \int_0^{2\pi} \int_{-\pi/2}^{\pi/2} \left[ 2(1+\nu)\frac{\partial^2 f}{\partial \overline{x} \partial \overline{y}} + \frac{\partial \overline{w}}{\partial \overline{x}}\frac{\partial \overline{w}}{\partial \overline{y}} \right] d\overline{x} d\overline{y}. \tag{18}$$

*3.3. Critical buckling analysis*

For critical buckling analysis, from Eqs. (A11), (A12), and (7), the resultant forces in the pre-buckling state can be obtained as

$$N_{x0} = \frac{\partial^2 F_0}{\partial y^2} = -\frac{P}{2\pi R}, \quad N_{xy0} = -\frac{\partial^2 F_0}{\partial x \partial y} = \frac{T}{2\pi R^2}, \quad N_{y0} = \frac{\partial^2 F_0}{\partial x^2} = 0. \tag{19}$$

Here, the subscript "0" denotes the quantities associated with the pre-buckling state. Using Eq. (19) and dropping the nonlinear terms in the governing equations (11) and (12), we have

$$c\overline{\nabla}^4 \overline{w} - \alpha \frac{\partial^2 f}{\partial \overline{x}^2} = -k_x \frac{\partial^2 \overline{w}}{\partial \overline{x}^2} + 2k_s \beta c \frac{\partial^2 \overline{w}}{\partial \overline{x} \partial \overline{y}}, \quad \overline{\nabla}^4 f + \alpha \frac{\partial^2 \overline{w}}{\partial \overline{x}^2} = 0, \tag{20}$$

By eliminating the stress function $f$ in Eq. (20), we can obtain the fourth-order Donnell equation as (Batdorf, 1947)

$$c\overline{\nabla}^4 \overline{w} + \alpha^2 \overline{\nabla}^{-4} \frac{\partial^4 \overline{w}}{\partial \overline{x}^4} + k_x \frac{\partial^2 \overline{w}}{\partial \overline{x}^2} - 2k_s \beta c \frac{\partial^2 \overline{w}}{\partial \overline{x} \partial \overline{y}} = 0. \tag{21}$$

Note that the operator $\nabla^{-4}$ is introduced to avoid using the eighth-order Donnell equation, as the higher-order derivatives in the Donnell equation may result in divergent trigonometric series for clamped boundary conditions (Batdorf, 1947), leading to inaccurate results. To illustrate this, a comparison of results based on the eighth-order Donnell equation and the fourth-order equation is provided in **Section S1** and **Table S1** in the Supplementary Materials. When the shell is clamped at both ends, the dimensionless transverse displacement $\bar{w}$ can be assumed as the following form (Yamaki, 1984),

$$\bar{w}(\bar{x},\bar{y}) = \sum_{m=1}^{\infty} a_m \left[ \phi_{m-1}(\bar{x},\bar{y}) + \phi_{m+1}(\bar{x},\bar{y}) \right], \tag{22}$$

with

$$\phi_m = \cos(m\bar{x}+\bar{y}) + (-1)^m \cos(m\bar{x}-\bar{y}), \tag{23}$$

where $a_m$ ($m = 1, 2, 3, \cdots$) are unknown coefficients. It can be proven that Eq. (22) satisfies the clamped boundary conditions Eq. (14) at both ends of the shell. Note that other forms of the solution $\bar{w}$ can also be used, and an alternative solution is provided in **Section S2** of the Supplementary Materials, which produces the same results as Eq. (22). Then, we apply the Galerkin method to Eq. (21), which leads to the condition

$$\int_0^{2\pi} \int_{-\pi/2}^{\pi/2} \mathcal{L}_0(\bar{w}) \left[ \phi_{j-1}(\bar{x},\bar{y}) + \phi_{j+1}(\bar{x},\bar{y}) \right] d\bar{x} d\bar{y} = 0, \quad j = 1, 2, 3, \cdots M, \tag{24}$$

where

$$\mathcal{L}_0(\bar{w}) = c\bar{\nabla}^4 \bar{w} + \alpha^2 \bar{\nabla}^{-4} \frac{\partial^4 \bar{w}}{\partial \bar{x}^4} + k_x \frac{\partial^2 \bar{w}}{\partial \bar{x}^2} - 2k_s \beta c \frac{\partial^2 \bar{w}}{\partial \bar{x} \partial \bar{y}}, \tag{25}$$

and $M$ represents the number of terms in the series that is used to approximate the exact solution.

Substituting Eq. (22) into Eq. (24) and performing integration, one has

$$\sum_{m=1}^{M} a_m \left[ \begin{array}{l} (P_{m-1}+Q_{m-1})A_{m-1,j} + (-1)^{m-1}(P_{m-1}-Q_{m-1})B_{m-1,j} \\ +(P_{m+1}+Q_{m+1})A_{m+1,j} + (-1)^{m+1}(P_{m+1}-Q_{m+1})B_{m+1,j} \end{array} \right] = 0, \tag{26}$$
$$j = 1, 2, \cdots, M,$$

with

$$P_m = c\left(m^2+\beta^2\right)^2 + \frac{\alpha^2 m^4}{\left(m^2+\beta^2\right)^2} - k_x m^2, \tag{27}$$

$$Q_m = 2k_s \beta c m, \tag{28}$$

$$A_{m,j} = \int_0^{2\pi} \int_{-\pi/2}^{\pi/2} \cos(m\bar{x}+\bar{y}) \left[ \phi_{j-1}(\bar{x},\bar{y}) + \phi_{j+1}(\bar{x},\bar{y}) \right] d\bar{x} d\bar{y}, \tag{29}$$

$$B_{m,j} = \int_0^{2\pi} \int_{-\pi/2}^{\pi/2} \cos(m\bar{x} - \bar{y}) \left[ \phi_{j-1}(\bar{x}, \bar{y}) + \phi_{j+1}(\bar{x}, \bar{y}) \right] d\bar{x} d\bar{y}, \tag{30}$$

In the calculation, the operator $\nabla^{-4}$ simply introduces the numerator expression obtained by applying $\nabla^4$ into the denominator of each term in the series.

Eq. (26) can be written into a matrix form as

$$\mathbf{C}[a_1 \ a_2 \ \cdots \ a_M]^{\mathrm{T}} = 0. \tag{31}$$

The coefficients $a_m$ ($m = 1, 2, \cdots, M$) have non-trivial solutions if and only if the determinant of $\mathbf{C}$ is equal to zero, i.e., $|\mathbf{C}| = 0$, from which the eigen equation can be obtained to determine the dimensionless critical buckling load $k_x^{\mathrm{cr}}$ or $k_s^{\mathrm{cr}}$ and the corresponding circumferential wavenumber $N_{\mathrm{cr}}$. It can be observed from Eqs. (26)-(31) that the dimensionless critical buckling load and the corresponding circumferential wavenumber for C-C cylindrical shells under combined torsional and axial loads only depend on the two geometric parameters $L/R$ and $R/h$ (from which $\alpha$ and $\beta$ are determined), and the Poisson's ratio $v$ (from which $c$ is determined), provided one of the external loads or a relationship between them is specified. The Wolfram Mathematica code for solving the eigenvalue problem is available on GitHub, with the link provided at the end of the paper.

*3.4. Post-buckling analysis*

For post-buckling analysis, the dimensionless transverse displacement $\bar{w}(\bar{x}, \bar{y})$ can be assumed in the following form (Yamaki, 1984):

$$\bar{w}(\bar{x}, \bar{y}) = \sum_{m=1}^{\infty} \sum_{n=0}^{\infty} a_{m,n} \left( \psi_{m-1,n} + \psi_{m+1,n} \right), \quad m = 1, 2, 3, \cdots, \quad n = 0, 1, 2 \cdots, \tag{32}$$

with

$$\psi_{m,n} = \cos(m\bar{x} + n\bar{y}) + (-1)^m \cos(m\bar{x} - n\bar{y}), \tag{33}$$

where $a_{m,n}$ are coefficients to be determined. It should be stated that in Yamaki (1984), Eq. (32) (32) was used to study the post-buckling of C-C cylindrical shells under pure torsion. Here, we demonstrate that this displacement form can also be applied to analyze the post-buckling of C-C cylindrical shells under combined torsional and axial loads. Substituting Eq. (32) into the compatibility condition Eq. (12) and rearranging the terms, we have

$$\bar{\nabla}^4 f = \sum_{p=0}^{\infty}\sum_{q=0}^{\infty} f_{pq}\psi_{p,q}, \quad p,q=0,1,2,\cdots; \quad f_{00}=0, \tag{34}$$

where

$$f_{pq} = \alpha p^2 (a_{p-1,q} + a_{p+1,q}) - \frac{1}{16}\beta^2 \{4 - 2\delta_{p0} - \delta_{q0}[3-(-1)^p]\}\sum_{m=1}^{\infty}\sum_{n=0}^{\infty} a_{m,n} A(p,q,m,n). \tag{35}$$

In Eq. (35), $\delta_{ij}$ is the Kronecker delta, and $A(p,q,m,n)$ is a linear combination of $a_{m,n}$, which is given by

$$A(p,q,m,n) = A_{p,q,m-1,n} + A_{p,q,m+1,n}, \tag{36}$$

$$\begin{aligned}
A_{p,q,r,s} &= a(p,q,r,s) + a(-p,-q,r,s) + a(p,q,-r,-s) \\
&\quad + (-1)^p \left[ a(p,-q,r,s) + a(-p,q,r,s) \right] \\
&\quad + (-1)^r \left[ a(p,q,r,-s) + a(p,q,-r,s) \right] \\
&\quad + (-1)^{p+r} \left[ a(p,-q,-r,s) + a(-p,q,r,-s) \right],
\end{aligned} \tag{37}$$

with

$$a(p,q,r,s) = (ps-qr)^2 (a_{p+r-1,q+s} + a_{p+r+1,q+s}). \tag{38}$$

Note that when the subscript $p \leq 0$ or $q < 0$, $a_{p,q} = 0$. The general solution of the dimensionless stress function $f$ can be expressed as (see **Section S3** in the Supplementary Materials for derivation details)

$$f(\bar{x},\bar{y}) = \frac{1}{2}\left\{ -\nu k_x + \sum_{p=0}^{\infty}\left[1+(-1)^p\right](-1)^{p/2} p^2 F_{p0} \right\} \bar{x}^2 \\
- \frac{k_x}{2\beta^2}\bar{y}^2 - \frac{ck_s}{\beta}\bar{x}\bar{y} + \sum_{p=0}^{\infty}\sum_{q=0}^{\infty} H_{pq}\psi_{p,q}, \tag{39}$$

where

$$H_{pq} = F_{pq} + G_{pq}, H_{p0} = F_{p0}, F_{pq} = \frac{f_{pq}}{\left(p^2+\beta^2 q^2\right)^2}, \quad F_{00} = 0,$$

$$G_{pq} = 2(2-\delta_{p0})(1-\delta_{q0})K_{pq}\sum_{k=0}^{\infty}\left[1+(-1)^{p+k}\right](-1)^{\frac{p+k}{2}}(k^2-\nu\beta^2 q^2)F_{kq}, \tag{40}$$

$$K_{pq} = \frac{p^2-\nu\beta^2 q^2}{(p^2+\beta^2 q^2)^2} \cdot \frac{(-1)^p \cosh(\pi\beta q)-1}{(1+\nu)\pi\beta q\left[(3-\nu)\sinh(\pi\beta q)-(-1)^p(1+\nu)\pi\beta q\right]}.$$

Substituting Eqs. (39) and (32) into the governing equation (11) and then applying the Galerkin method, one has

$$\int_0^{2\pi} \int_{-\pi/2}^{\pi/2} \mathcal{L}(\bar{w}, \bar{f})\left(\psi_{r-1,s} + \psi_{r+1,s}\right) d\bar{x} d\bar{y} = 0, \quad r = 1,2,3,\cdots, \quad s = 0,1,2,\cdots, \quad (41)$$

with

$$\mathcal{L}(\bar{w}, \bar{f}) = c\bar{\nabla}^4 \bar{w} - \alpha \frac{\partial^2 f}{\partial \bar{x}^2} - \beta^2 \left( \frac{\partial^2 \bar{f}}{\partial \bar{y}^2} \frac{\partial^2 \bar{w}}{\partial \bar{x}^2} - 2 \frac{\partial^2 \bar{w}}{\partial \bar{x} \partial \bar{y}} \frac{\partial^2 \bar{f}}{\partial \bar{x} \partial \bar{y}} + \frac{\partial^2 \bar{f}}{\partial \bar{x}^2} \frac{\partial^2 \bar{w}}{\partial \bar{y}^2} \right). \quad (42)$$

Substituting the expressions of $w$ and $f$ into Eq. (41) and performing integration, the following algebraic equations in terms of $a_{m,n}$ can be obtained

$$\left[1 - (-1)^r \delta_{s0}\right]\left\{L_{r-1,s}\left[a_{r-2,s} + (1+\delta_{r1})a_{r,s}\right] + L_{r+1,s}(a_{r,s} + a_{r+2,s}) + \alpha\left[(r-1)^2 H_{r-1,s} + (r+1)^2 H_{r+1,s}\right]\right\}$$

$$+ \frac{64}{\pi} \beta c s k_s \sum_{m=1} N_{m,r} a_{m,s} + \left\{\beta^2 s^2 \left[a_{r-2,s} + (2+\delta_{r1})a_{r,s} + a_{r+2,s}\right] - 2\alpha \delta_{r1} \delta_{s0}\right\}\left\{\sum_{p=0}\left[1 + (-1)^p\right](-1)^{p/2} p^2 F_{p0} - \nu k_x\right\}$$

$$-(1+\delta_{s0})\left[2(r^2+1)a_{r,s} + (r-1)^2 a_{r-2,s} + (r+1)^2 a_{r+2,s}\right] k_x - \frac{1}{2}\beta^2 \sum_{p=0}\sum_{q=0} H_{pq} A(p,q,r,s) = 0,$$

$$r = 1,2,3,\cdots, \quad s = 0,1,2,\cdots,$$

(43)

in which

$$L_{r,s} = c(r^2 + \beta^2 s^2)^2, \quad (44)$$

$$N_{m,r} = \frac{\left[1 - (-1)^{m+r}\right](-1)^{(m-r+1)/2} mr(m^2 + r^2 - 2)}{(m^2 - r^2)\left[(m-r)^2 - 4\right]\left[(m+r)^2 - 4\right]}. \quad (45)$$

Note that when $k_x = 0$, Eq. (43) reduces to the result derived by Yamaki (1984) for pure torsion. Based on Eq. (43), the coefficients $a_{m,n}$ can be determined when the values of Poisson's ratio $\nu$, geometric parameter $\alpha$, wavenumber $N$, and the applied torque $k_s$ and compressive force $k_x$ are prescribed. Additionally, from Eqs. (35) to (37) and Eq. (40), one can find that $H_{pq}$ is a second-order algebraic expression in terms of $a_{m,n}$, and $A(p, q, r, s)$ is a linear combination of $a_{m,n}$. Therefore, the coupled terms of $H_{pq}$ and $A(p, q, r, s)$ in Eq. (43) leads to a system of cubic algebraic equations in terms of $a_{m,n}$, which can be solved by various numerical iteration methods. In this work, we solve these equations using the *FindRoot* function in commercial software Wolfram

Mathematica with a Newton iteration solver. The code is available on GitHub with the link provided at the end of the paper.

Moreover, by using Eqs. (32) and (39), the dimensionless end shortening and twisting angle of the shell can be calculated by

$$\bar{\delta} = (1-v^2)\frac{k_x}{\alpha} + \frac{v}{\alpha}\sum_{p=0}\left[1+(-1)^p\right](-1)^{p/2} p^2 F_{p0} + \frac{1}{2\alpha}\sum_{m=1}\sum_{n=0}a_{m,n}\left[1-(-1)^m \delta_{n0}\right]$$
$$\times \left[(m-1)^2(a_{m-2,n}+a_{m,n}) + (m+1)^2(a_{m,n}+a_{m+2,n})\right], \tag{46}$$

$$\bar{\varphi} = \frac{2}{\alpha}(1+v)ck_s - \frac{32\beta}{\pi\alpha}\sum_{k=1}\sum_{m=1}\sum_{n=1}nN_{k,m}a_{k,n}a_{m,n}. \tag{47}$$

## 4. Finite element simulations

Finite element simulations for the buckling and post-buckling behavior of cylindrical shells under combined torsional and axial loads are performed in the commercial software Abaqus 2021 (Dassault Systèmes, France). The shell is modeled using a 3D shell model with a fixed radius of $R$ = 100 mm. The thickness and length vary depending on the radius-to-thickness ratio $R/h$ and length-to-radius ratio $L/R$ considered in the numerical examples. The Young's modulus and Poisson's ratio of the shell are taken as $E$ = 3.5 GPa and $v$ = 0.3, respectively. In the simulation, the S4R (4-node stress/displacement shell element with reduced integration) linear element is used. Boundary conditions and external loads are applied to reference points, which are constrained to the edges of the shell through rigid body constraints. For the C-C boundary conditions considered in simulations, the bottom end of the shell is fixed, while the top end, where external loads are applied, is free to translate and rotate about the longitudinal direction. Note that these boundary conditions are slightly different from the experimental setup for post-buckling under compression with pre-torsion. In the experiments, a pre-twist angle is initially applied and then held constant during compressive loading and unloading.

The critical buckling load and critical wavenumber can be determined by the eigenvalue buckling analysis, in which a linear perturbation is applied to trigger the buckling. Specifically, a small, concentrated force is applied for buckling under pure compression while a small twisting moment is employed for buckling under pure torsion. In particular, for the buckling under combined compression and torsion studied in **Table 1** and **Fig. 5**, a small force and twisting moment are both applied as linear perturbations. For buckling under compression with pre-torsion, or torsion with pre-tension/compression, which are required in the corresponding post-buckling simulations shown in **Figs. 6-11**, the preload is first applied before introducing the corresponding linear perturbation. The lowest eigenvalue represents the critical buckling load, and the corresponding buckling eigenmode gives the critical wavenumber. The obtained buckling eigenmodes are then utilized in the post-buckling analysis as a small geometric imperfection to trigger the bifurcation modes. The imperfection profile is defined as a linear superposition of the first three buckling eigenmodes, with the amplitudes of the three modes set to 1%, 0.5% and 0.25% of the shell thickness, respectively. Then, a displacement-controlled quasi-static loading method is used to determine the post-buckling equilibrium path, and a small damping factor of $10^{-8}$ is introduced to stabilize the simulation. Abaqus input files for selected examples are available on GitHub with the link provided at the end of the paper.

## 5. Results and discussion

In this section, numerical results for the buckling and post-buckling analyses of C-C thin cylindrical shells under combined torsional and axial loads predicted by FEA and the theoretical model solved using the Galerkin method are both presented. The effect of combined loads, radius-to-thickness ratio, and length-to-radius ratio on the critical buckling loads, critical circumferential wavenumbers, buckling patterns, and post-buckling equilibrium paths are investigated in detail. For the critical buckling analysis, we mainly focus on the case of cylindrical shells under combined torsion and compression. For the post-buckling analysis, we consider all five loading methods used in the experiments: pure compression, pure torsion, compression with pre-torsion, torsion with pre-tension, and torsion with pre-compression, while the boundary condition at the loaded end is free to rotate and translate along the shell's longitudinal direction for all considered cases in the theoretical model and FEA (as noted previously, the boundary condition in the experiments for post-buckling under compression with pre-torsion is different, where the pre-twist is applied before compression and then kept constant during compressive loading and unloading processes). Moreover, unless otherwise stated, 5 terms (i.e., $M = 5$) are retained in the series solution of the Galerkin method for the critical buckling analysis, and 38 terms are retained for the post-buckling analysis (a convergence study is presented in **Section S4** and **Fig. S1** in the Supplementary Materials). Further, due to space limitations, "Galerkin method" is abbreviated as "GM" in the legends of **Figs. 6-11**.

**Table 1.** Dimensionless critical buckling loads of C-C cylindrical shells under different types of loads ($L/R = 1$ and $v = 0.3$)

| Loading method | Dimensionless critical buckling load | Solution method | $R/h = 200$ | $R/h = 300$ | $R/h = 400$ | $R/h = 500$ |
|---|---|---|---|---|---|---|
| Pure compression ($k_s = 0$) | $k_x^{cr}$ | FEA | 12.3927 (12)* | 18.6683 (15) | 24.7153 (18) | 30.9620 (20) |
| | | Galerkin method | 12.2711 (11) | 18.4216 (15) | 24.5734 (18) | 30.8092 (20) |
| Pure torsion ($k_x = 0$) | $k_s^{cr}$ | FEA | 47.8120 (13) | 64.2357 (14) | 79.5049 (16) | 93.8624 (17) |
| | | Galerkin method | 47.7030 (12) | 64.0856 (14) | 79.2719 (16) | 93.6708 (17) |
| Combined torsion and compression ($k_s = 2k_x$) | $k_x^{cr}$ | FEA | 8.6942 (12) | 12.6487 (15) | 16.4554 (16) | 20.2279 (18) |
| | | Galerkin method | 8.6233 (12) | 12.5488 (14) | 16.3214 (16) | 20.0181 (18) |

*The numbers in parentheses are the corresponding circumferential wavenumber

*5.1. Buckling under combined torsion and compression*

To begin with, the numerical results obtained by the Galerkin method is validated by comparing with FEA results in **Table 1**, where $k_x^{cr}$ is the dimensionless critical buckling compressive force, and $k_s^{cr}$ is the dimensionless critical buckling torque. It can be observed that under different types of loads, the critical buckling loads and the corresponding circumferential wavenumbers of C-C cylindrical shells with various radius-to-thickness ratios acquired by the Galerkin method are in good agreement with those obtained from FEA simulations. The minor differences in the circumferential wavenumber for $R/h = 200$ and 300 may result from the effect of shear deformation. It should be noted that the Galerkin method not only shows high accuracy, but also significantly higher efficiency compared to FEA simulation. To illustrate this, **Table 2** compares the computational time of the Galerkin method and FEA simulation for buckling analysis of cylindrical shells, using buckling under pure torsion for different length-to-radius ratios as an example. Both calculations are performed on a desktop computer equipped with an Intel® Xeon® E5-1650 v3 CPU (6 cores, 12 threads, 3.5 GHz). It is seen that the Galerkin method requires only a few seconds, whereas the FEA simulation takes several minutes to achieve the same level of accuracy. Convergence studies for both methods are provided in **Tables S2** and **S3** in the Supplementary Material. These results indicate that the Galerkin method is a highly efficient and accurate approach for the buckling analysis of cylindrical shells.

**Table 2.** Comparison of the computational time $t$ (s) of Galerkin method and FEA for critical buckling of C-C cylindrical shells with $R/h = 400$ under pure torsion

|  | Solution method | $L/R = 1$ | $L/R = 2$ | $L/R = 3$ | $L/R = 4$ |
|---|---|---|---|---|---|
| $k_s^{cr}$ | FEA | 79.5049 (16)* | 225.193 (12) | 415.225 (10) | 638.847 (9) |
|  | Galerkin method | 79.2719 (16) | 224.293 (12) | 415.576 (10) | 640.381 (9) |
| $t$ (s) | FEA | 145 | 78 | 88 | 76 |
|  | Galerkin method | 1.0693 | 2.5057 | 2.5096 | 2.5228 |

*The numbers in parentheses are the corresponding circumferential wavenumber

Then, the critical buckling behavior of C-C cylindrical shells under combined torsion and compression is studied based on the theoretical model integrated with the Galerkin method. To identify how the critical buckling behavior is affected by the combined loads, we assume a ratio between the shear stress and axial stress $\tau/\sigma$ induced by torsion and compression. **Fig. 5(a)** and **(b)** illustrate the critical axial strain $\varepsilon_{cr}$ as a function of $\tau/\sigma$ and the radius-to-thickness ratio $R/h$ (with a fixed $L/R = 1$) or length-to-radius ratio $L/R$ (with a fixed $R/h = 400$), respectively. Here, 7 terms (i.e., $M = 7$) are retained in the series solution when using the Galerkin method for cylindrical shells with $L/R$ greater than 1. It is shown that the critical axial strain decreases as $\tau/\sigma$ increases, which means that shear stress tends to reduce the critical axial strain for cylindrical shells under combined torsion and compression. Moreover, with a given $L/R$ (or $R/h$), the smaller $R/h$ (or $L/R$), the higher the critical axial strain. Note that under pure compression (i.e., $\tau/\sigma = 0$), the theoretical model predicts that the critical axial strain remains nearly unchanged as $L/R$ increases. A similar phenomenon was observed in Yamaki (1984). This is because the Donnell shell model assumes that the in-plane displacements are negligibly small, whereas they become significant in long cylindrical shells.

The critical circumferential wavenumber as a function of $\tau/\sigma$ and $R/h$ (with a fixed $L/R = 1$) or $L/R$ (with a fixed $R/h = 400$) are depicted in **Fig. 5(c)** and **(d)**, respectively. For a given $L/R$ and $\tau/\sigma$, the critical circumferential wavenumber increases as $R/h$ increases, which means that thinner shells (with larger $R/h$) have a smaller buckling wavelength. Conversely, for a given $R/h$ and $\tau/\sigma$, the critical circumferential wavenumber decreases as $L/R$ increases, indicating that longer shells (with larger $L/R$) have a larger buckling wavelength. Additionally, for a given $R/h$ and $L/R$, an increase in $\tau/\sigma$ tends to reduce the critical circumferential wavenumber, and this effect becomes more pronounced for shells with larger $R/h$ and $L/R$. To better illustrate this, **Fig. 5(e)** presents the critical buckling patterns of C-C cylindrical shells with $R/h = 400$ and $L/R = 1$ or $L/R = 2$ under combined torsion and compression for various $\tau/\sigma$. It is seen that as $\tau/\sigma$ increases from 0 to 1, the critical circumferential wavenumber decreases from 18 to 16 for the shell with $R/h = 400$ and $L/R = 1$, whereas for the shell with $R/h = 400$ and $L/R = 2$, it decreases from 17 to 12. Therefore, the buckling wavenumber and wavelength of cylindrical shells can be effectively tuned by introducing pre-torsion or pre-compression.

On the other hand, when $\tau/\sigma = 0$, corresponding to pure compression, the shell exhibits a diamond-shaped pattern. In contrast, when combined torsion and compression are applied, the shell buckles into a diagonal-shaped pattern even when the applied torsion is very small (e.g., $\tau/\sigma = 0.005$ and $0.01$). A comparison of the critical buckling patterns predicted by the theoretical model and FEA simulations is provided in **Fig. S2** in the Supplementary Materials, and a good agreement can be observed. These results indicate that the critical buckling patterns of thin cylindrical shells are highly sensitive to shear stress. Even a small shear stress (compared to the axial stress induced by compression) can cause the critical buckling pattern under compression to turn from a diamond-shaped pattern to a diagonal-shaped pattern.

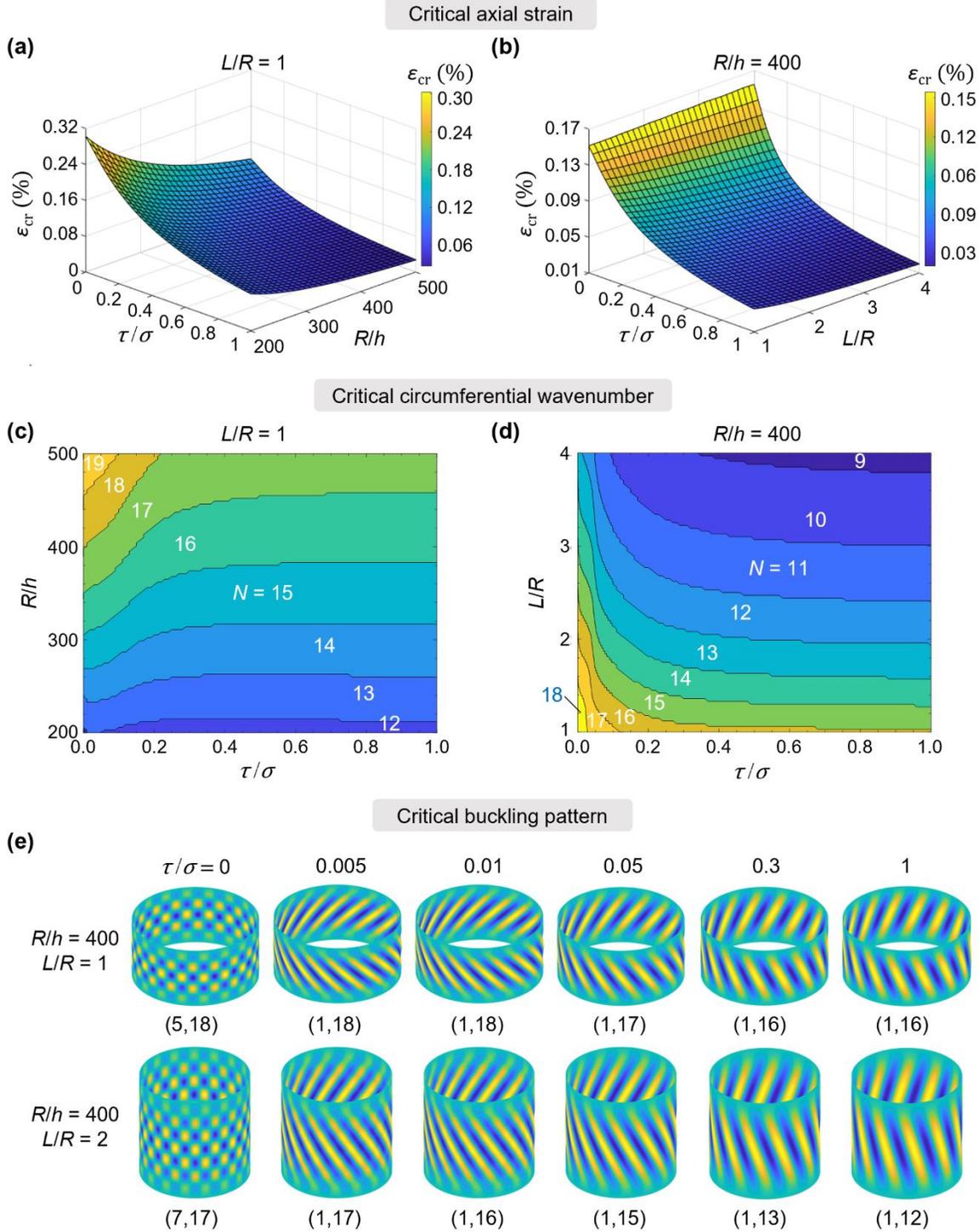

**Fig. 5.** Critical buckling of C-C cylindrical shells under combined torsion and compression. (a) Critical axial strain $\varepsilon_{cr}$ as a function of $\tau/\sigma$ and $R/h$ with a fixed $L/R = 1$. (b) Critical axial strain $\varepsilon_{cr}$ as a function of $\tau/\sigma$ and $L/R$ with a fixed $R/h = 400$. (c) Contour plot of the critical circumferential wavenumber $N_{cr}$ as a function of $\tau/\sigma$ and $R/h$ with a fixed $L/R = 1$. (d) Contour plot of the critical circumferential wavenumber $N_{cr}$ as a function of $\tau/\sigma$ and $L/R$ with a fixed $R/h = 400$. (e) Critical buckling patterns for different $\tau/\sigma$ and $L/R$ with a fixed $R/h = 400$. The two numbers in parentheses represent the axial half-wavenumber and the circumferential wavenumber, respectively.

### 5.2. Post-buckling under compression with pre-torsion

Next, we study the post-buckling behavior of C-C cylindrical shells under compression with pre-torsion. Four different values of pre-torsion are considered, i.e., $T_0 = 0$, $0.25T_{cr0}$, $0.50T_{cr0}$, and $0.75T_{cr0}$, with $T_{cr0}$ being the critical buckling torque under pure torsion. As mentioned in **Section 2**, the normalized compressive force $P/P_{cr0}$ will be used to present the numerical results for post-buckling under compression. For convenience, we define $\Sigma = P/P_{cr0}$. **Fig. 6(a)-(d)** shows the normalized compressive force $\Sigma$ versus the dimensionless end shortening $\bar{\delta}$ of a C-C cylindrical shell with $R/h = 200$ and $L/R = 1$ during loading and unloading, as predicted by FEA and the theoretical model using the Galerkin method. The dimensionless torque $k_{s0}$ for pre-torsion in the four cases are taken as 0, 11.95, 23.90, and 35.85, respectively, and the resulting critical buckling loads are 100%, 80.9%, 56.1%, and 28.6% of the critical load under pure compression. This demonstrates that the critical buckling load of a cylindrical shell under compression can be adjusted by applying pre-torsion. Due to the small geometric imperfection introduced in FEA simulations, the critical buckling loads predicted by FEA, corresponding to the peaks of the curves, are slightly lower than those of the theoretical model. After the onset of buckling, both FEA and the theoretical model predict an unstable post-buckling equilibrium path, indicated by a dramatic decrease in the buckling load. This suggests that snapping occurs after buckling, along with a significantly reduced load-carrying capability of the shell. Moreover, the post-buckling equilibrium paths predicted by the theoretical model agree well with the simulation results for different cases except for when snapping occurs, as a force-controlled loading method is used in the Galerkin method.

During the loading process, when the pre-torsion is absent (i.e., $T_0 = 0$ in **Fig. 6(a)**) or relatively small (e.g., $T_0 = 0.25T_{cr0}$ in **Fig. 6(b)**), the shell snaps from one buckling mode to another, and the circumferential wavenumber decreases one by one in the post-buckling regime, as observed in the experiment of **Fig. 2(b)**. Particularly, the shell consistently exhibits a diamond-shaped pattern under pure compression, and the wavelength of the pattern increases after each snapping (see the patterns at points $A_1$ to $D_1$ shown at the bottom of **Fig. 6(a)**). However, the shell shows a twisted diamond-shaped pattern when a small pre-torsion is applied. Under compressive loading, the twisted pattern gradually merges in a diagonal manner (see the patterns at points $A_1$ to $D_1$ shown at the bottom of **Fig. 6(b)**). As will be demonstrated in **Fig. 7(a)**, this twisted diamond-shaped pattern eventually transitions into a diagonal-shaped pattern under further loading. In contrast, when the applied pre-torsion is relatively large (e.g., $T_0 = 0.50T_{cr0}$ and $0.75T_{cr0}$ in **Fig. 6(c)** and **(d)**), the shell directly snaps into a diagonal-shaped pattern with $N = 12$ from the critical buckling state under compression, and no further snapping occurs with continued loading, during which the compressive force gradually decreases until reaching a plateau.

On the other hand, when the pre-torsion is zero or relatively small, the shell follows a new equilibrium path with lower buckling load during unloading, as illustrated by the dashed lines in **Fig. 6(a)** and **(b)**. Note that in the theoretical analysis, we can identify various bifurcation branches for different $N$, but we cannot determine which branch corresponds to the unloading path. Therefore, only unloading results from FEA are indicated by dashed lines. During the unloading process, the shell displays a diamond-shaped pattern with the axial half-wavenumber decreased from 2 to 1 when $T_0 = 0$. For $T_0 = 0.25T_{cr0}$, the shell exhibits a twisted diamond-shaped pattern, which tends to merge to form a diagonal-shaped pattern. In other words, in compression-dominated post-buckling, the shell follows an irreversible loading/unloading equilibrium path. In contrast, when the pre-torsion is relatively large (**Fig. 6(c)** and **(d)**), the shell exhibits torsion-dominated post-buckling behavior and follows the loading equilibrium path during unloading (except for the snapping part after buckling). It should be stated that because a specified pre-twist angle between the two ends of the shell is held constant throughout the test to serve as pre-torsion in experiments, the buckling pattern evolution and post-buckling equilibrium path under compression with pre-torsion in FEA (which permits longitudinal rotation at the loaded end) differ from the experimental results. However, the FEA results still qualitatively capture several key experimental observations shown in **Fig. 2**, such as the shell showing a compression-dominated post-buckling behavior when the pre-torsion is relatively small, and the shell exhibiting a twisted diamond-shaped pattern when the pre-torsion is relatively large. This indicates that pre-torsion has a significant influence on the post-buckling behavior of cylindrical shells under compression, affecting both the post-buckling equilibrium path as well as the buckling pattern evolution.

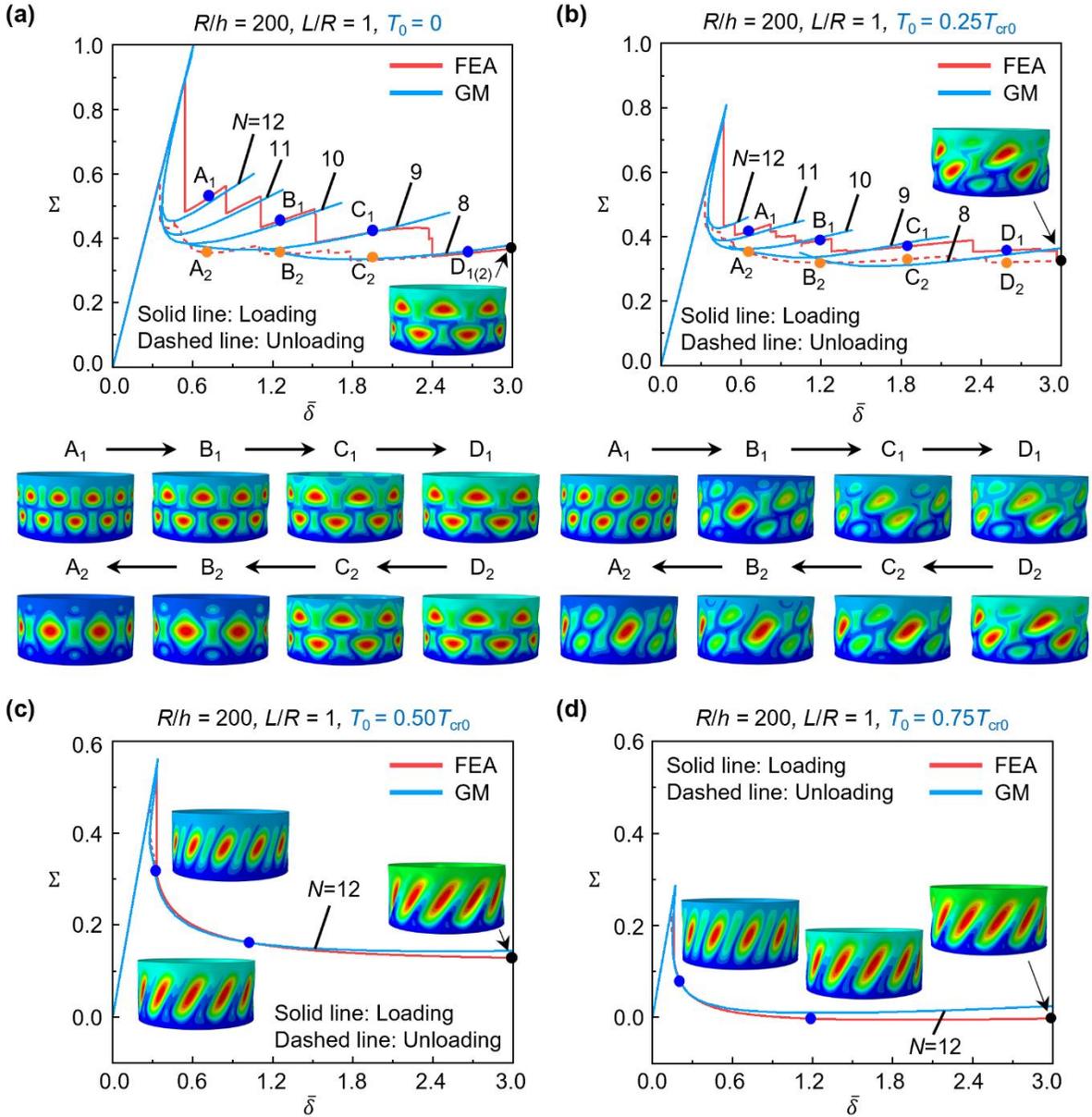

**Fig. 6.** Normalized compressive force Σ versus the dimensionless end shortening $\bar{\delta}$ for post-buckling behavior of C-C cylindrical shells with $R/h = 200$ and $L/R = 1$ under compression with different values of pre-torsion: (a) $T_0 = 0$, (b) $T_0 = 0.25T_{cr0}$, (c) $T_0 = 0.50T_{cr0}$, and (d) $T_0 = 0.75T_{cr0}$, where $T_0$ is the applied pre-torsion and $T_{cr0}$ is the critical buckling torque under pure torsion. The buckling patterns are obtained from FEA. In the Galerkin method, the normalized critical compressive force and circumferential wavenumbers ($\Sigma_{cr}$, $N_{cr}$) are (1, 11), (0.8089, 12), (0.5605, 12), (0.2863, 12), respectively.

Subsequently, the effect of radius-to-thickness ratio on the post-buckling behavior of C-C cylindrical shells under compression with pre-torsion is examined based on the theoretical model and FEA. **Fig. 7(a)-(d)** shows the post-buckling equilibrium paths and buckling pattern evolutions with $T_0 = 0.25T_{cr0}$ for $R/h$ = 200, 300, 400, and 500, respectively. $L/R$ is fixed at 1. For the four shells, the dimensionless torques $k_{s0}$ for pre-torsion are taken as 11.95, 16.06, 19.85, and 23.46, respectively, and the resulting critical buckling loads are all approximately 80% of the critical load under pure compression. Moreover, the circumferential wavenumbers at the critical buckling states predicted by the theoretical model are 12, 15, 17, and 19, respectively. After the onset of buckling, FEA shows that the shells snap into post-buckling modes with circumferential numbers of 12, 13, 14, and 15, respectively. In the post-buckling regime, the shell snaps from one buckling mode to another and the circumferential wavenumber decreases one by one, similar to the case of pure compression (see **Fig. S3** in the Supplementary Materials for post-buckling behavior of C-C cylindrical shells with various $R/h$ under pure compression). As $R/h$ increases, more snapping processes (in the FEA results) and bifurcation branches (in the theoretical results) can be observed during deformation to the same dimensionless end shortening. Additionally, during the loading process, the shell transitions from a twisted diamond-shaped pattern to a diagonal-shaped pattern with $N$ = 10, 12, 12, and 14, predicted by both the theoretical model and FEA, for $R/h$ = 200, 300, 400, and 500, respectively (see the patterns at points $A_1$ to $D_1$ at the bottom of **Fig. 7(a)-(d)**). The results with larger pre-torsion $T_0 = 0.50T_{cr0}$ and $0.75T_{cr0}$ for different $R/h$ are provided in **Figs. S4** and **S5**, respectively, in the Supplementary Materials. When the applied pre-torsion is $T_0 = 0.50T_{cr0}$ (see **Fig. S4** in the Supplementary Materials), the shell snaps into a diagonal-shaped pattern after critical buckling occurs, with no further snapping under continued loading for $R/h$ = 200 and 300. However, for $R/h$ = 400 and 500, the shell first snaps into a twisted diamond-shaped pattern before snapping to a diagonal-shaped pattern during loading. When the applied pre-torsion is $T_0 = 0.75T_{cr0}$ (see **Fig. S5** in the Supplementary Materials), the shell directly transitions into the diagonal-shaped pattern after buckling for different $R/h$.

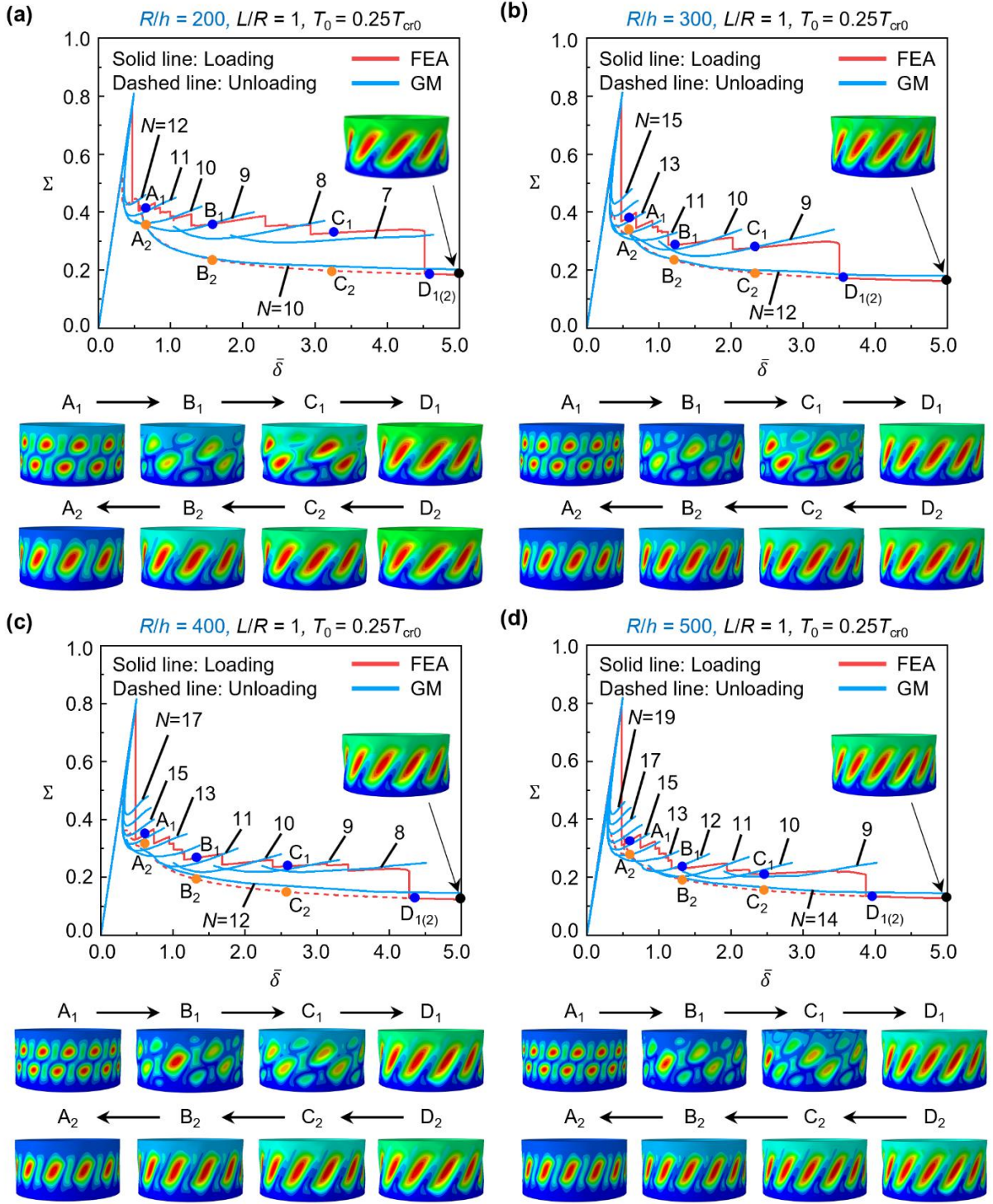

**Fig. 7.** Normalized compressive force $\Sigma$ versus the dimensionless end shortening $\bar{\delta}$ for post-buckling behavior of C-C cylindrical shells with $L/R = 1$ under compression with pre-torsion $T_0 = 0.25T_{cr0}$: (a) $R/h = 200$, (b) $R/h = 300$, (c) $R/h = 400$, and (d) $R/h = 500$. The buckling patterns are obtained from FEA. In the Galerkin method, the normalized critical compressive force and circumferential wavenumbers ($\Sigma_{cr}$, $N_{cr}$) are (0.8089, 12), (0.8124, 15), (0.8155, 17), and (0.8183, 19), respectively.

Further, the effect of length-to-radius ratio on the post-buckling behavior of C-C cylindrical shells under compression with pre-torsion is investigated in **Fig. 8(a)-(d)**. $R/h$ is fixed at 200 and $L/R$ varies from 0.75, 1, 1.5 to 2. The pre-torsion is considered as $T_0 = 0.25T_{cr0}$, leading to a decrease in the critical buckling load by about 20% compared to that under pure compression. In the four cases, the dimensionless torques $k_{s0}$ for pre-torsion are taken as 8.00, 11.95, 21.48, and 32.99, respectively, and the corresponding circumferential wavenumbers when critical buckling occurs predicted by the theoretical model using the Galerkin method are 13, 12, 12, and 11, respectively. It should be noted that as $L/R$ increases, more series terms are required in the Galerkin method to achieve a convergent solution. However, this significantly increases the computational cost. Therefore, here we mainly focus on relatively short shells, while FEA results for slightly longer shells (e.g., $L/R$ = 3 and 4) are provided in **Fig. S6** in the Supplementary Materials. When the shell is very short (e.g., $L/R$ = 0.75), FEA shows that it directly snaps into a diagonal-shaped pattern with $N$ = 11 after the onset of buckling under compression with pre-torsion. However, when the shell is relatively long (e.g., $L/R$ = 1, 1.5, and 2), the shell exhibits a compression-dominated post-buckling behavior (see **Fig. S7** in the Supplementary Materials for the post-buckling of C-C cylindrical shells with various $L/R$ under pure compression), snapping from one buckling mode to another as the circumferential wavenumber reduces one by one. During the loading process, the shell transitions from a twisted diamond-shaped pattern to a diagonal-shaped pattern with $N$ = 10, 9, and 8, predicted by both the theoretical model and FEA, for $L/R$ = 1, 1.5, and 2, respectively. Note that for cylindrical shells with $L/R$ smaller than 1, it is also possible to show such compression-dominated post-buckling behavior when the applied pre-torsion is even smaller. For example, as shown in **Fig. S8** in the Supplementary Materials, the cylindrical shell with $R/h$ = 200 and $L/R$ = 0.75 morphs from a twisted diamond-shaped pattern into a diagonal-shaped pattern under compression with a pre-torsion $T_0$=0.10$T_{cr0}$. Results of the post-buckling behavior of C-C cylindrical shells under compression with a relatively large pre-torsion $T_0$ = 0.50$T_{cr0}$ or 0.75$T_{cr0}$ for different $L/R$ are provided in **Figs. S9** and **S10**, respectively, in the Supplementary Materials. Except for the cases of $L/R$ = 1.5 and 2 with $T_0$ = 0.50$T_{cr0}$ (**Fig. S9(c)** and **(d)**), all other cases show a torsion-dominated post-buckling behavior, and the shell directly snaps into a diagonal-shaped pattern after buckling.

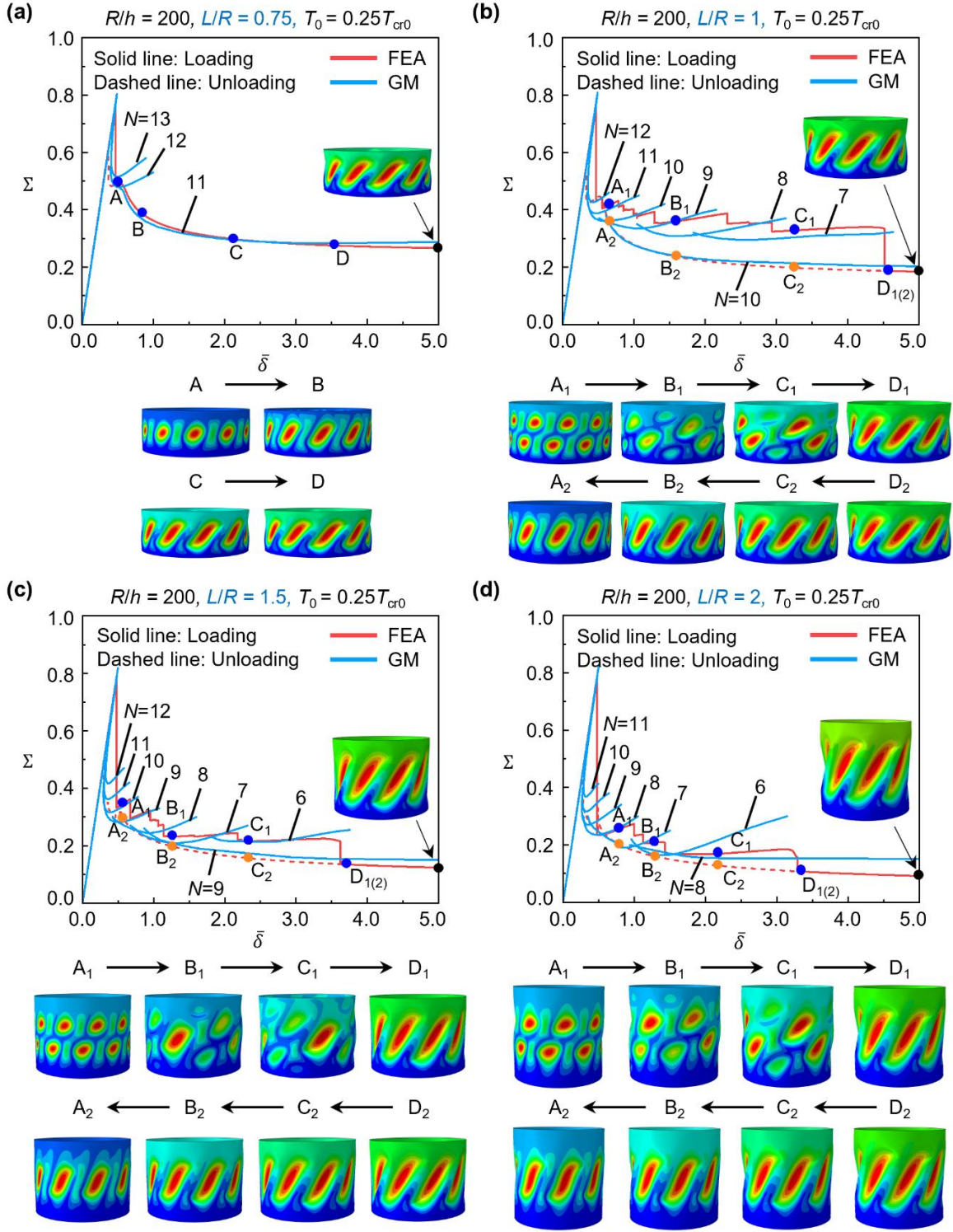

**Fig. 8.** Normalized compressive force $\Sigma$ versus the dimensionless end shortening $\bar{\delta}$ for post-buckling behavior of C-C cylindrical shells with $R/h = 200$ under compression with pre-torsion $T_0 = 0.25T_{cr0}$: (a) $L/R$

= 0.75, (b) $L/R = 1$, (c) $L/R = 1.5$, and (d) $L/R = 2$. The buckling patterns are obtained from FEA. In the Galerkin method, the normalized critical compressive force and circumferential wavenumbers ($\Sigma_{cr}$, $N_{cr}$) are (0.8037, 13), (0.8089, 12), (0.8187, 12), and (0.8257, 11), respectively.

*5.3. Post-buckling under torsion with pre-tension/compression*

Having studied the post-buckling behavior of cylindrical shells under compression with pre-torsion, in this subsection, we investigate how pre-tension/compression affect the post-buckling behavior of cylindrical shells under torsion. **Fig. 9(a)** presents the dimensionless torque $k_s$ versus the dimensionless twisting angle $\bar{\varphi}$ for post-buckling of a C-C cylindrical shell with $R/h$ = 400 and $L/R = 1$ under torsion with different values of pre-tension (i.e., $P_0 = 0$, $-2P_{cr0}$, $-4P_{cr0}$, and $-8P_{cr0}$, where $P_{cr0}$ is the classical value under pure compression given in Eq. (1)). The dimensionless axial forces $k_{x0}$ for pre-tension are taken as 0, −49.06, −98.12, and −196.24, respectively, leading to an increase in the critical buckling torque by 0%, 150%, 280%, and 510%. It is seen that both FEA and the theoretical model predict a smooth post-buckling equilibrium path for various values of pre-tension, meaning that no snapping occurs after the onset of buckling. In particular, under pure torsion, the post-buckling equilibrium path is unstable, as evidenced by a decreasing post-buckling load, which signifies a reduced load-carrying capacity of the buckled shell. However, when a sufficient pre-tension is applied (e.g., $P_0 = -2P_{cr0}$, $-4P_{cr0}$, and $-8P_{cr0}$), the post-buckling equilibrium path becomes stable, characterized by an increasing post-buckling load, indicating that the shell can carry more torsional load after buckling. This demonstrates that the axial pre-tension can not only increase the critical buckling load under torsion but also alter the stability of the post-buckling equilibrium path. The corresponding buckling pattern evolutions are shown at the bottom of **Fig. 9(a)**. It is seen that the shell always exhibits a diagonal-shaped pattern under torsion with pre-tension. During the loading process, the circumferential wavenumber remains unchanged, while the deflection of the buckling pattern increases. Additionally, pre-torsion tends to increase the circumferential wavenumber, thereby decreasing the wavelength of the buckling pattern. For instance, the circumferential wavenumber increases from 16 to 22 as the pre-tension varies from 0 to $-8P_{cr0}$. This conclusion matches the experimental observations shown in **Figs. 1** and **3(b)**.

The post-buckling equilibrium paths and post-buckling patterns of C-C cylindrical shells under torsion with different values of pre-compression are plotted in **Fig. 9(b)**. Here, the pre-compression is considered to be 0, $0.1P_{cr0}$, $0.25P_{cr0}$, and $0.50P_{cr0}$, respectively (the corresponding dimensionless axial forces $k_{x0}$ are 0, 2.45, 6.13, and 12.26), resulting in a decrease in the critical buckling load under torsion by 0%, 8.3%, 20.9%, and 43.1%. When the pre-compression is relatively small (e.g., $P_0 = 0.1P_{cr0}$), the shell exhibits post-buckling behavior similar to the pure torsion case, with the torque gradually decreasing as the twisting angle increases after critical buckling occurs. As the pre-compression increases to $P_0 = 0.25P_{cr0}$, the theoretical model predicts a smooth post-buckling equilibrium path without snapping, while FEA shows that snapping occurs during the loading process and the shell buckles into a diagonal-shaped pattern with a reduced circumferential wavenumber. This discrepancy may be induced by the small geometric imperfection considered in the FEA simulation. When the pre-compression is relatively large (e.g., $P_0 = 0.5P_{cr0}$), both FEA and the theoretical model indicate that snapping occurs followed by the critical buckling, while distinct snapping behavior is predicted due to the different methods used to find the equilibrium path. In the theoretical model solved by the Galerkin method (a torque-controlled loading method is used), both the torque and twisting angle decrease to zero after buckling, meaning that the shell snaps back to the initial state. In contrast, in the FEA (a displacement-controlled loading method is used), the torque suddenly drops to zero and the shell transitions from a diagonal-shaped pattern to a twisted diamond-shaped pattern, as observed in the experiment of **Fig. 3(d)**. It should be noted that the torque variation obtained from these two methods also differs from that observed in experiments, although all approaches capture the key feature that torque tends to decrease to zero after buckling under torsion with a relatively large pre-compression. This discrepancy is likely due to differences in deformation modes between the experiments and the FEA/theoretical model. In experiments, we observe that the buckling pattern in the circumferential direction under torsion does not develop simultaneously and is not perfectly periodic. For instance, as shown in pattern A of **Fig. 3(d)**, the front part of the shell exhibits significantly larger deformation than the remaining parts after the onset of buckling. This may be caused by imperfect boundary constraints or minor geometric imperfections in the shell. Conversely, in our FEA and theoretical model, the circumferential buckling pattern appears simultaneously and remains perfectly periodic under torsion. As a result, the measured torque variation in experiments follows a different trend compared to that predicted by FEA and theoretical model.

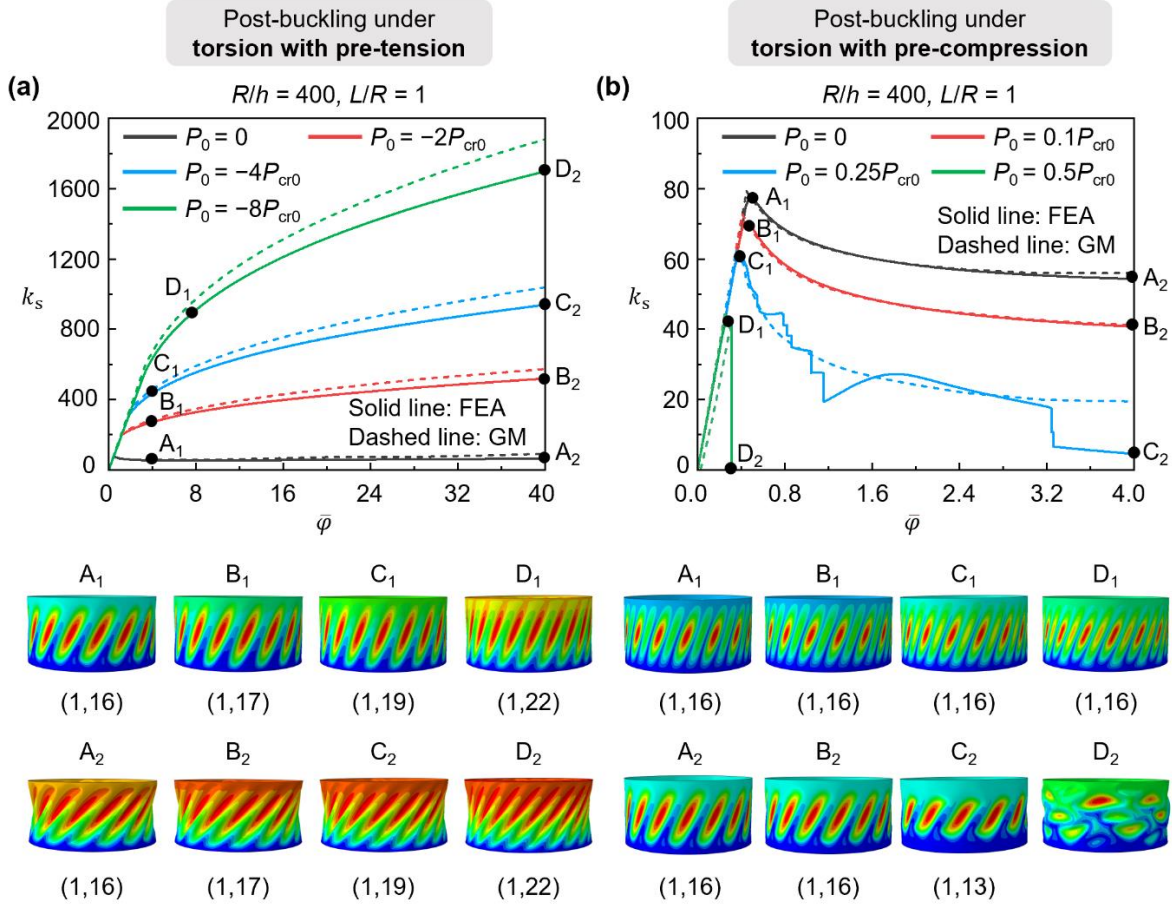

**Fig. 9.** Dimensionless torque $k_s$ versus the dimensionless twisting angle $\bar{\varphi}$ for post-buckling behavior of C-C cylindrical shells with $R/h = 400$ and $L/R = 1$ under torsion with different values of (a) pre-tension or (b) pre-compression. The buckling patterns are obtained from FEA. The numbers in parentheses represent the axial half-wavenumber and circumferential wavenumber. In the Galerkin method, the dimensionless critical buckling torques and the corresponding circumferential wavenumbers ($k_s^{cr}$, $N_{cr}$) are (79.27, 16), (197.71, 17), (301.31, 19), (483.98, 22) in Fig. (a), and (79.27, 16), (72.72, 16), (62.68, 16), and (45.13, 16) in Fig. (b).

The effect of radius-to-thickness ratio on the post-buckling of C-C cylindrical shells under torsion with pre-tension or pre-compression is investigated in **Fig. 10**. Here, $L/R$ is set to 1, and $R/h$ varies from 200, 300, 400 to 500. **Fig. 10(a)** illustrates the post-buckling equilibrium paths and post-buckling patterns for various $R/h$ with a pre-tension $P_0 = -0.25P_{cr0}$. For the four cases, the dimensionless axial forces $k_{x0}$ for pre-tension are taken as −3.07, −4.60, −6.13, and −7.67, respectively, leading to an increase in the critical buckling torque by approximately 20% compared to the case under pure torsion. It is shown that with a relatively small pre-tension, the post-buckling load first decreases and then increases after the onset of buckling, similar to the case of pure torsion (see **Fig. S11** in the Supplementary Materials). For different $R/h$, no snapping occurs, and the wavenumber does not change during loading when a pre-tension is applied. When the shell is subjected to torsion with pre-compression $P_0 = 0.25P_{cr0}$ (see **Fig. 10(b)**, and the dimensionless axial forces $k_{x0}$ for pre-compression are taken as 3.07, 4.60, 6.13, and 7.67, respectively), the critical buckling torque is decreased by about 20% compared to that under pure torsion. In this case, the theoretical model predicts a smooth post-buckling equilibrium path for various $R/h$. However, for shells with $R/h = 400$ and 500, snapping is observed during the loading process in FEA. As mentioned before, this discrepancy may be due to the small geometric imperfection introduced in the FEA simulations. Additional results of post-buckling of C-C cylindrical shells under torsion with larger pre-tension $P_0 = -P_{cr0}$ or pre-compression $P_0 = 0.5P_{cr0}$ for different $R/h$ are provided in **Fig. S12** in the Supplementary Materials. Similar to the observations in **Fig. 9**, when the pre-tension is relatively large, the torque continuously goes up after buckling and no snapping occurs during the entire loading process. When the pre-compression is relatively large, the torque suddenly drops to zero and the shell snaps into a twisted diamond-shaped pattern.

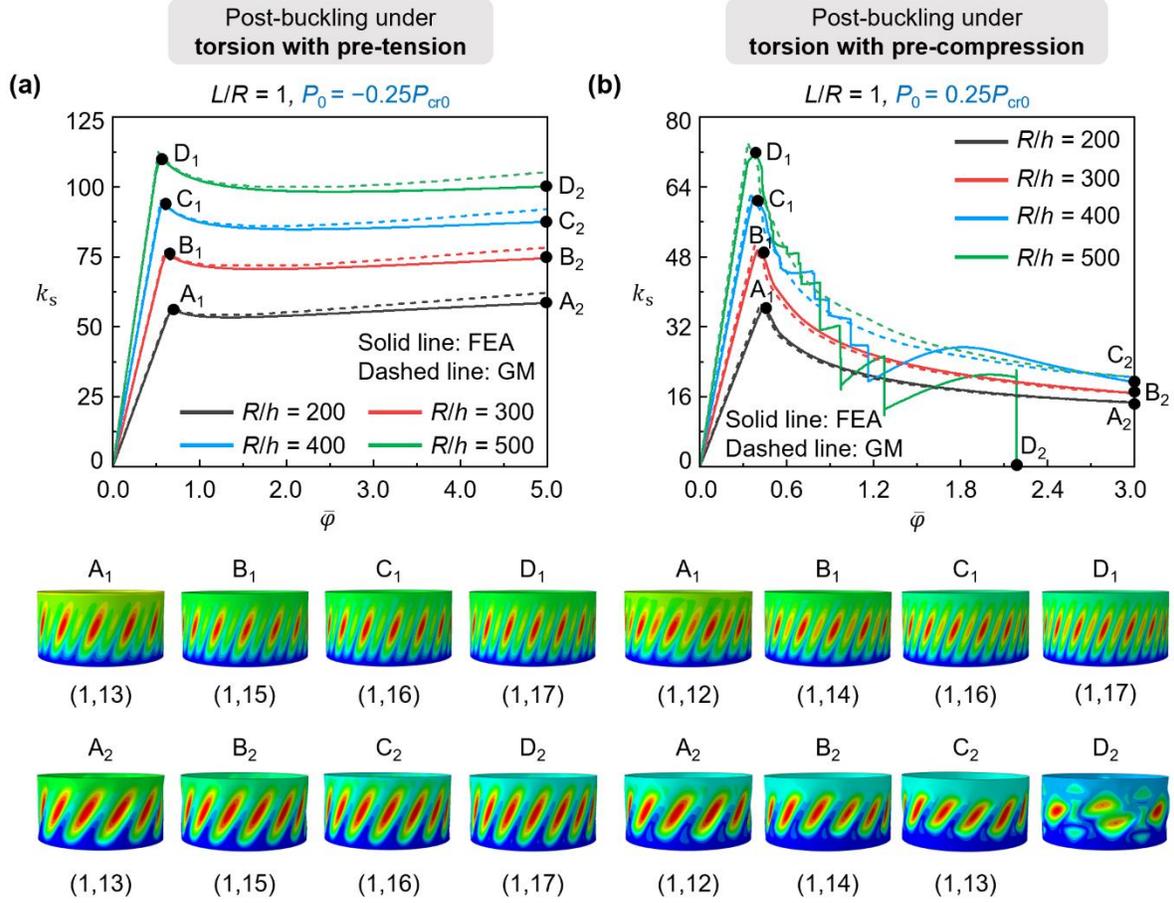

**Fig. 10.** Dimensionless torque $k_s$ versus the dimensionless twisting angle $\bar{\varphi}$ for post-buckling behavior of C-C cylindrical shells with $L/R = 1$ under torsion with (a) pre-tension $P_0 = -0.25P_{cr0}$ or (b) pre-compression $P_0 = 0.25P_{cr0}$. $R/h$ varies from 200, 300, 400 to 500. The buckling patterns are obtained from FEA. The numbers in parentheses represent the axial half-wavenumber and circumferential wavenumber. In the Galerkin method, the dimensionless critical buckling torques and the corresponding circumferential wavenumbers ($k_s^{cr}$, $N_{cr}$) are (57.45, 13), (77.23, 15), (95.26, 16), and (112.47, 17) in Fig. (a), and (37.38, 12), (50.47, 14), (62.68, 16), and (74.19, 17) in Fig. (b).

Finally, we examine the effect of length-to-radius ratio on the post-buckling behavior of cylindrical shells under torsion with pre-tension or pre-compression in **Fig. 11**. Here, $R/h$ is fixed at 200, and $L/R$ varies from 1, 2, 3, to 4. The pre-tension and pre-compression are considered as $P_0 = -0.25P_{cr0}$ and $0.25P_{cr0}$, respectively, which leads to an increase or a decrease in the critical buckling torque by approximately 20% compared to that under pure torsion. For the four shells, the dimensionless axial forces $k_{x0}$ for pre-tension are −3.07, −12.26, −27.60, and −49.06, and for pre-compression are 3.07, 12.26, 27.60, and 49.06, respectively. As shown in **Fig. 11(a)**, for the various $L/R$ considered, the torque first increases linearly with the twisting angle up to the critical buckling state, then gradually decreases, and eventually begins to increase as the twisting angle continues to increase. During the loading process, no snapping occurs, and the shell exhibits a diagonal-shaped pattern, with the circumferential wavenumber decreasing as $L/R$ increases. In contrast, both FEA and the theoretical model predict that the shell with pre-compression snaps with further loading after buckling except in the case of $L/R = 1$ (**Fig. 11(b)**), and FEA shows that the shell transitions into a twisted diamond-shaped pattern. This indicates that with a given $R/h$, the post-buckling behavior of longer shells under torsion is more sensitive to the effect of pre-compression. Additional results for the post-buckling behavior of C-C cylindrical shells with different $L/R$ under torsion with pre-tension $P_0 = -P_{cr0}$ or pre-compression $P_0 = 0.5P_{cr0}$ are presented in **Fig. S13** in the Supplementary Materials. As expected, results show that the torque continuously increases without snapping after buckling when the shell is subjected to torsion with a larger pre-tension, while snapping occurs after the onset of buckling when the shell is under torsion with a larger pre-compression.

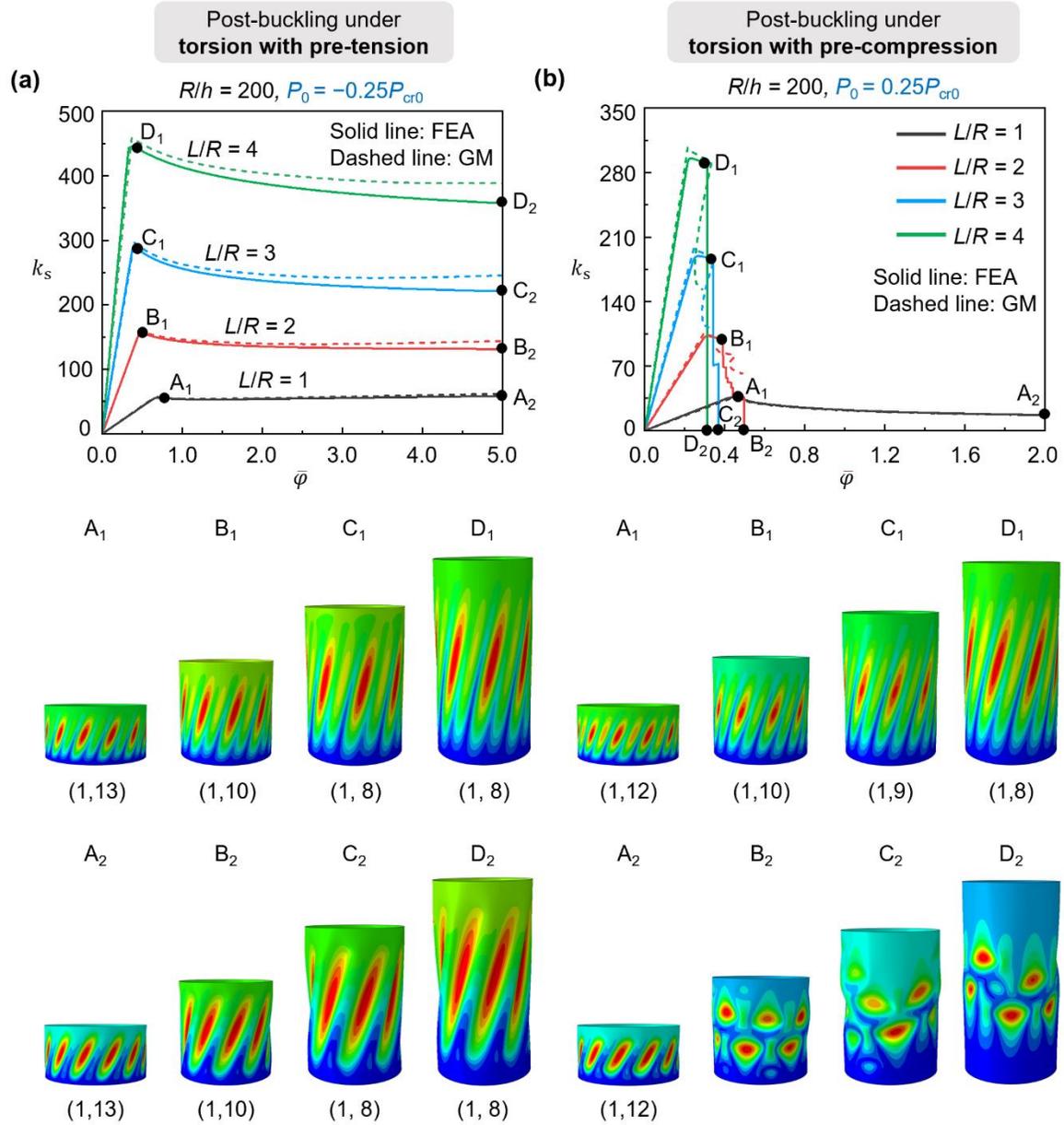

**Fig. 11.** Dimensionless torque $k_s$ versus the dimensionless twisting angle $\bar{\varphi}$ for post-buckling behavior of C-C cylindrical shells with $R/h = 200$ under torsion with (a) pre-tension $P_0 = -0.25P_{cr0}$ or (b) pre-compression $P_0 = 0.25P_{cr0}$. $L/R$ varies from 1, 2, 3 to 4. The buckling patterns are obtained from FEA. The numbers in parentheses represent the axial half-wavenumber and circumferential wavenumber. In the Galerkin method, the dimensionless critical buckling torques and the corresponding circumferential wavenumbers ($k_s^{cr}$, $N_{cr}$) are (57.45, 13), (159.83, 10), (296.50, 8), and (458.44, 8) in Fig. (a), and (37.38, 12), (106.14, 10), (198.15, 9), and (306.79, 8) in Fig. (b).

# 6. Conclusions

In summary, we have studied the buckling and post-buckling behavior of cylindrical shells under combined torsional and axial loads through a combination of experiments, theoretical modeling, and finite element simulations. Three types of combined loads were considered for thin cylindrical shells with clamped-clamped ends: compression with pre-torsion, torsion with pre-tension, and torsion with pre-compression. Also, post-buckling under pure compression or pure torsion was studied for comparison purposes. Critical buckling and post-buckling analyses were performed using the Donnell shell theory integrated with the Galerkin method, from which the critical buckling load, critical circumferential wavenumber, buckling pattern, and the post-buckling equilibrium path of cylindrical shells under a single load or combined loads could be determined. The theoretical predictions were validated through finite element simulations, which also qualitatively captured the experimental observations. The main conclusions for buckling and post-buckling of thin cylindrical shells with clamped-clamped ends (the loaded end is free to rotate and translate along the longitudinal direction) under combined loads are summarized as follows.

- The critical buckling pattern of cylindrical shells under combined torsion and compression is highly sensitive to shear stress. Even if the shear stress induced by the applied torsion is very small compared to the axial stress caused by compression, the shell may exhibit a diagonal-shaped torsional buckling pattern under combined torsion and compression.
- For post-buckling of cylindrical shells under compression with pre-torsion, when the pre-torsion is relatively small (e.g., 25% of the critical buckling load under pure torsion), the shell snaps from one buckling mode to another and the circumferential wavenumber decreases by one during each snapping process. In this case, the shell transitions from a twisted diamond-shaped pattern to a diagonal-shaped pattern under sufficiently large deformation. However, when the pre-torsion is relatively large (e.g., 75% of the critical buckling load under pure torsion), the shell directly snaps into the diagonal-shaped pattern after the onset of buckling and no snapping occurs with further loading.
- For post-buckling of cylindrical shells under torsion with pre-tension, when the pre-tension is relatively small (e.g., 25% of the critical buckling load under pure compression), the torque first decreases and then increases with the twisting angle after critical buckling occurs. However, when the pre-tension is relatively large (e.g., 100% of the critical buckling load under pure compression), the torque continuously increases after buckling. In both cases, no snapping occurs, and the shell always exhibits a diagonal-shaped pattern with the circumferential wavenumber unchanged during the loading process.

- For post-buckling of cylindrical shells under torsion with pre-compression, when the pre-compression is relatively small (e.g., 10% of the critical buckling load under pure compression), the torque gradually decreases as the twisting angle increases after critical buckling occurs and the shell morphs into a diagonal-shaped pattern. However, when the pre-compression is relatively large (e.g., 50% of the critical buckling load under pure compression), snapping occurs after buckling and the shell transitions from a diagonal-shaped pattern to a twisted diamond-shaped pattern.

It should be noted that the above conclusions are based on our numerical studies for thin and short cylindrical shells (the radius-to-thickness ratios considered are between 200 and 500, and the length-to-radius ratios considered are between 0.75 and 4), where the predictions of the Donnell shell theory show good agreements with finite element simulations. For the buckling and post-buckling behavior of thicker and longer cylindrical shells under combined torsional and axial loads, factors such as shear deformation, coupled bending and stretching deformation, and edge effects may significantly influence their mechanical behavior. In such cases, more advanced shell theories will be required.

Our work reveals the role of torsion-compression/tension coupling in the buckling instabilities of cylindrical shells, which could provide potential guidelines for the design of functional cylindrical shell structures. Also, we envision that the various buckling patterns presented in this work can guide the discovery of cylindrical shell buckling-inspired foldable structures such as origami systems driven by combined loads.

## Data availability

The Wolfram Mathematica codes and Abaqus input files used to generate the numerical results in this paper are available on GitHub: https://github.com/Stanford-ZhaoLab/Shell-buckling

# Acknowledgements

This work was supported by the National Science Foundation Career Award CMMI-2145601 and National Science Foundation Award CPS 2201344.

# Appendix. Derivation of the governing equations

Based on the Donnell shell theory, the non-zero strain components in a thin cylindrical shell can be written as

$$\varepsilon_x = \varepsilon_x^0 + z\kappa_x, \quad \varepsilon_y = \varepsilon_y^0 + z\kappa_y, \quad \gamma_{xy} = \gamma_{xy}^0 + z\kappa_{xy}, \tag{A1}$$

with

$$\varepsilon_x^0 = \frac{\partial u}{\partial x} + \frac{1}{2}\left(\frac{\partial w}{\partial x}\right)^2, \quad \varepsilon_y^0 = \frac{\partial v}{\partial y} - \frac{w}{R} + \frac{1}{2}\left(\frac{\partial w}{\partial y}\right)^2, \quad \gamma_{xy}^0 = \frac{\partial u}{\partial y} + \frac{\partial v}{\partial x} + \frac{\partial w}{\partial x}\frac{\partial w}{\partial y}, \tag{A2}$$

$$\kappa_x = -\frac{\partial^2 w}{\partial x^2}, \quad \kappa_y = -\frac{\partial^2 w}{\partial y^2}, \quad \kappa_z = -2\frac{\partial^2 w}{\partial x \partial y}, \tag{A3}$$

where $\varepsilon_x^0$, $\varepsilon_y^0$ and $\gamma_{xy}^0$ are the strain components on the middle plane of the shell, $\kappa_x$, $\kappa_y$, and $\kappa_{xy}$ are the curvature components, and $u(x, y)$, $v(x, y)$ and $w(x, y)$ are the displacement components along $x$, $y$ and $z$ axes, respectively.

The stress-strain relations are given by

$$\sigma_x = \frac{E}{1-v^2}(\varepsilon_x + v\varepsilon_y), \quad \sigma_y = \frac{E}{1-v^2}(\varepsilon_y + v\varepsilon_x), \quad \sigma_{xy} = \frac{E}{2(1+v)}\gamma_{xy}, \tag{A4}$$

where $E$ is the Young's modulus, and $v$ is the Poisson's ratio.

Then, the resultant internal forces and moments can be defined as

$$\{N_x, N_y, N_{xy}\} = \int_{-h/2}^{h/2} \{\sigma_x, \sigma_y, \sigma_{xy}\} dz, \tag{A5}$$

$$\{M_x, M_y, M_{xy}\} = \int_{-h/2}^{h/2} \{\sigma_x, \sigma_y, \sigma_{xy}\} z dz. \tag{A6}$$

Substituting Eqs. (A1) and (A4) into Eqs. (A5) and (A6), one can obtain that

$$N_x = \frac{Eh}{1-v^2}\left(\varepsilon_x^0 + v\varepsilon_y^0\right), \quad N_y = \frac{Eh}{1-v^2}\left(v\varepsilon_x^0 + \varepsilon_y^0\right), \quad N_{xy} = \frac{Eh}{2(1+v)}\gamma_{xy}^0, \tag{A7}$$

$$M_x = D(\kappa_x + \nu\kappa_y), \; M_y = D(\nu\kappa_x + \kappa_y), \; M_{xy} = \frac{D(1-\nu)}{2}\kappa_{xy}, \tag{A8}$$

where $D = Eh^3/[12(1-\nu^2)]$ is the flexural rigidity of the shell.

Moreover, the variation of the strain energy of the cylindrical shell can be written as

$$\begin{aligned}\delta U &= \int_{-h/2}^{h/2}\int_0^{2\pi R}\int_{-L/2}^{L/2}(\sigma_x\delta\varepsilon_x + \sigma_y\delta\varepsilon_y + \sigma_{xy}\delta\gamma_{xy})dxdydz \\ &= \int_0^{2\pi R}\int_{-L/2}^{L/2}(N_x\delta\varepsilon_x^0 + N_y\delta\varepsilon_y^0 + N_{xy}\delta\gamma_{xy}^0 + M_x\delta\kappa_x + M_y\delta\kappa_y + M_{xy}\delta\kappa_{xy})dxdy.\end{aligned} \tag{A9}$$

The variation of the work done by external work is written as

$$\delta W = -\int_0^{2\pi R}\int_{-L/2}^{L/2}(f_x\delta u + f_y\delta v + f_z\delta w)dxdy, \tag{A10}$$

where $f_x$, $f_y$, and $f_z$ are the body forces along $x$, $y$, and $z$ axes, respectively. By using the principle of minimum potential, i.e., $\delta U + \delta W = 0$, the equilibrium equations of thin cylindrical shells can be obtained as

$$\frac{\partial N_x}{\partial x} + \frac{\partial N_{xy}}{\partial y} + f_x = 0, \tag{A11}$$

$$\frac{\partial N_y}{\partial y} + \frac{\partial N_{xy}}{\partial x} + f_y = 0, \tag{A12}$$

$$\begin{aligned}&\frac{\partial^2 M_x}{\partial x^2} + 2\frac{\partial^2 M_{xy}}{\partial x \partial y} + \frac{\partial^2 M_y}{\partial y^2} + \frac{N_y}{R} + \frac{\partial}{\partial x}\left(N_x\frac{\partial w}{\partial x} + N_{xy}\frac{\partial w}{\partial y}\right) \\ &+ \frac{\partial}{\partial y}\left(N_y\frac{\partial w}{\partial y} + N_{xy}\frac{\partial w}{\partial x}\right) + f_z = 0,\end{aligned} \tag{A13}$$

and the corresponding boundary conditions at $x = \pm L/2$ are given by

$$N_x = 0 \text{ or } u = 0, \tag{A14}$$

$$N_{xy} = 0 \text{ or } v = 0, \tag{A15}$$

$$N_x\frac{\partial w}{\partial x} + N_{xy}\frac{\partial w}{\partial y} + \frac{\partial M_x}{\partial x} + \frac{\partial M_{xy}}{\partial y} = 0 \text{ or } w = 0, \tag{A16}$$

$$M_x = 0 \text{ or } \frac{\partial w}{\partial x} = 0, \tag{A17}$$

$$M_{xy} = 0 \text{ or } \frac{\partial w}{\partial y} = 0. \tag{A18}$$

The equilibrium equations can be simplified by introducing the Airy's stress function $F(x,y)$, which is defined as

$$N_x = \frac{\partial^2 F}{\partial y^2}, \ N_y = \frac{\partial^2 F}{\partial x^2}, \ N_{xy} = -\frac{\partial^2 F}{\partial x \partial y}. \tag{A19}$$

Moreover, from Eqs. (A2) and (A7), we have

$$\frac{\partial^2 F}{\partial y^2} - \nu \frac{\partial^2 F}{\partial x^2} = N_x - \nu N_y = Eh\left[\frac{\partial u}{\partial x} + \frac{1}{2}\left(\frac{\partial w}{\partial x}\right)^2\right], \tag{A20}$$

$$\frac{\partial^2 F}{\partial x^2} - \nu \frac{\partial^2 F}{\partial y^2} = N_y - \nu N_x = Eh\left[\frac{\partial v}{\partial y} - \frac{w}{R} + \frac{1}{2}\left(\frac{\partial w}{\partial y}\right)^2\right], \tag{A21}$$

$$-2(1+\nu)\frac{\partial^2 F}{\partial x \partial y} = 2(1+\nu)N_{xy} = Eh\left(\frac{\partial u}{\partial y} + \frac{\partial v}{\partial x} + \frac{\partial w}{\partial x}\frac{\partial w}{\partial y}\right). \tag{A22}$$

By eliminating the displacement components $u$ and $v$ in Eqs. (A20) to (A22), the compatibility equation can be obtained as

$$\nabla^4 F + \frac{Eh}{R}\frac{\partial^2 w}{\partial x^2} = Eh\left[\left(\frac{\partial^2 w}{\partial x \partial y}\right)^2 - \frac{\partial^2 w}{\partial x^2}\frac{\partial^2 w}{\partial y^2}\right], \tag{A23}$$

where

$$\nabla^4(\bullet) = \frac{\partial^4(\bullet)}{\partial x^4} + 2\frac{\partial^4(\bullet)}{\partial x^2 \partial y^2} + \frac{\partial^4(\bullet)}{\partial y^4}. \tag{A24}$$

On the other hand, in the absence of body forces, i.e., $f_x = f_y = f_z = 0$, the Airy's stress function satisfies the equilibrium equations (A11) and (A12) automatically. Then, substituting Eqs. (A8) and (A19) into Eq. (A13), the equilibrium equations in terms of the Airy's stress function $F$ and the transverse displacement $w$ can be written as

$$D\nabla^4 w - \frac{1}{R}\frac{\partial^2 F}{\partial x^2} = \frac{\partial^2 F}{\partial y^2}\frac{\partial^2 w}{\partial x^2} - 2\frac{\partial^2 F}{\partial x \partial y}\frac{\partial^2 w}{\partial x \partial y} + \frac{\partial^2 F}{\partial x^2}\frac{\partial^2 w}{\partial y^2}. \tag{A25}$$

Eqs. (A23) and (A25) are the two governing equations for thin cylindrical shells based on the Donnell shell theory, which only contains two unknown variables, i.e., the Airy's stress function $F(x, y)$ and the transverse displacement $w(x, y)$.

The cylindrical shell is clamped at both ends and subjected to a compressive force $P$ and a torque $T$, and the boundary conditions can be written as

$$w = \frac{\partial w}{\partial x} = \frac{\partial^2 u}{\partial y^2} = \frac{\partial v}{\partial y} = 0, \text{ at } x = \pm \frac{L}{2}, \tag{A26}$$

$$\int_0^{2\pi R} N_x dy = -P, \quad R\int_0^{2\pi R} N_{xy} dy = T. \tag{A27}$$

From Eqs. (A21) and (A22), one can obtain that

$$\frac{\partial^2 u}{\partial y^2} = -\frac{1}{R}\frac{\partial w}{\partial x} - \frac{\partial w}{\partial x}\frac{\partial^2 w}{\partial y^2} - \frac{1}{Eh}\left[\frac{\partial^3 F}{\partial x^3} + (2+\nu)\frac{\partial^3 F}{\partial x \partial y^2}\right], \tag{A28}$$

$$\frac{\partial v}{\partial y} = \frac{1}{Eh}\left(\frac{\partial^2 F}{\partial x^2} - \nu\frac{\partial^2 F}{\partial y^2}\right) + \frac{w}{R} - \frac{1}{2}\left(\frac{\partial w}{\partial y}\right)^2. \tag{A29}$$

At the two clamped ends, $w = \partial w/\partial x = 0$. In Eq. (A18), we specify $\partial w/\partial y = 0$. Then, the latter two boundary conditions in terms of $u$ and $v$ in Eq. (A26) can be expressed by the stress function $F(x, y)$ as

$$\frac{\partial^3 F}{\partial x^3} + (2+\nu)\frac{\partial^3 F}{\partial x \partial y^2} = 0, \quad \frac{\partial^2 F}{\partial x^2} - \nu\frac{\partial^2 F}{\partial y^2} = 0. \tag{A30}$$

Additionally, the end shortening can be obtained as

$$\begin{aligned}\Delta &= -\frac{1}{2\pi R}\int_0^{2\pi R}\int_{-L/2}^{L/2}\frac{\partial u}{\partial x}dxdy \\ &= -\frac{1}{2\pi R}\int_0^{2\pi R}\int_{-L/2}^{L/2}\left[\frac{1}{Eh}\left(\frac{\partial^2 F}{\partial y^2} - \nu\frac{\partial^2 F}{\partial x^2}\right) - \frac{1}{2}\left(\frac{\partial w}{\partial x}\right)^2\right]dxdy.\end{aligned} \tag{A31}$$

The twisting angle is given by

$$\begin{aligned}\varphi &= \frac{1}{2\pi R^2}\int_0^{2\pi R}\int_{-L/2}^{L/2}\left(\frac{\partial u}{\partial y} + \frac{\partial v}{\partial x}\right)dxdy \\ &= -\frac{1}{2\pi R^2}\int_0^{2\pi R}\int_{-L/2}^{L/2}\left(\frac{2(1+\nu)}{Eh}\frac{\partial^2 F}{\partial x \partial y} + \frac{\partial w}{\partial x}\frac{\partial w}{\partial y}\right)dxdy.\end{aligned} \tag{A32}$$